\documentclass[twocolumn]{aastex63}
\usepackage{amsmath}
\usepackage{xcolor}

\def\ms{\mbox{$M_{\ast}$}}
\def\msun{\mbox{M$_{\odot}$}}

\def\sfr{\mbox{${\rm SFR}$}}
\def\ssfr{\mbox{${\rm sSFR}$}}
\def\sfruv{\mbox{SFR$_{\rm FUV}$}}
\def\sfrlir{\mbox{SFR$_{\rm IR}$}}
\def\sfrtot{\mbox{SFR$_{\rm FUV+IR}$}}
\def\ssfrtot{\mbox{sSFR$_{\rm FUV+IR}$}}
\def\kfuv{\mbox{$\mathcal{K}_{\rm FUV}$}}
\def\kir{\mbox{$\mathcal{K}_{\rm IR}$}}
\def\mobsc{\mbox{$M_{\ast,\rm obsc}$}}
\submitjournal{ApJ}

\shorttitle{FUV+IR star formation rates since $z\sim4$}
\shortauthors{Aldo Rodr\'iguez-Puebla et al. }
\graphicspath{{./}{figures/}}

\begin{document}

\title{The star-forming main sequence and the contribution of dust-obscured star formation since $z\sim4$ from the FUV+IR luminosity functions}

\correspondingauthor{Aldo Rodr\'iguez-Puebla}
\email{apuebla@astro.unam.mx}

\author[0000-0002-0170-5358]{Aldo Rodr\'iguez-Puebla}
\affiliation{Instituto de Astronom\'ia
Universidad Nacional Aut\'onoma de M\'exico, 
A. P. 70-264, 04510, M\'exico, D.F., M\'exico}
\author[0000-0002-3461-2342]{Vladimir Avila-Reese}
\affiliation{Instituto de Astronom\'ia
Universidad Nacional Aut\'onoma de M\'exico, 
A. P. 70-264, 04510, M\'exico, D.F., M\'exico}
\author[0000-0001-9553-8230]{Mariana Cano-D\'iaz}
\affiliation{CONACYT Research fellow - Instituto de Astronom\'ia
Universidad Nacional Aut\'onoma de M\'exico, 
A. P. 70-264, 04510, M\'exico, D.F., M\'exico}
\author{S. M. Faber}
\affiliation{UCO/Lick Observatory, Department of Astronomy and Astrophysics, University of California, Santa Cruz, CA 95064, USA}
\author{Joel R. Primack}
\affiliation{Physics Department, University of California, Santa Cruz, CA 95064, USA}
\author{Jos\'e Franco}
\affiliation{Instituto de Astronom\'ia
Universidad Nacional Aut\'onoma de M\'exico, 
A. P. 70-264, 04510, M\'exico, D.F., M\'exico}
\author{I. Aretxaga}
\affiliation{Instituto Nacional de Astrof\'isica, \'Optica y Electr\'onica (INAOE), Aptdo. Postal 51 y 216, 72000 Puebla, Pue., M\'exico}
\author{Eder Santiago-Mayoral}
\affiliation{Instituto de Astronom\'ia
Universidad Nacional Aut\'onoma de M\'exico, 
A. P. 70-264, 04510, M\'exico, D.F., M\'exico}



\begin{abstract}
An analytical approach is proposed to study the evolution of the star-forming galaxy (SFG) 
main sequence (MS) and the fraction of dust-obscured SF up to $z\sim4$. 
Far-ultraviolet (FUV) and infrared (IR) star formation rates, SFRs, are described as 
conditional probability functions of \ms. We convolve them with the galaxy stellar 
mass function (GSMF) of SFGs to derive the FUV and IR LFs. The 2 SF 
modes formalism is used to describe starburst galaxies. By fitting observed FUV and IR LFs, 
the parametrization of $\sfruv-\ms$ and $\sfrlir-\ms$ are constrained. 
Our derived $\sfrtot-\ms$ reproduces the evolution of the MS as compared 
to other observational inferences. At any redshift, we find that the \ssfrtot--\ms\ 
relation for MS SFGs approaches to a power law at the high-mass end. At lower masses,
it bends and eventually the slope sign changes from negative to positive at very low masses. 
At $z\sim0$, this change of sign is at $\ms\sim5\times10^{8}\msun$ 
close to dust-obscured SF regime, 
$\ms\sim6\times10^{8}\msun$. The slope sign change is related to the knee of the FUV LF. 
Our derived dust-obscured fractions agree with previous determinations 
at $0\leq z\leq2.5$. Dust-obscured fractions
depend strongly on mass with almost no dependence with redshift at $z\gtrsim1.2$.
At $z\lesssim0.75$ high-mass galaxies become more ``transparent"  
compared to their high redshift counterparts. On the opposite, 
low- and intermediate-mass galaxies have become more obscured by dust. 
The joint evolution of the GSMF and the FUV and IR LFs is a
promising approach to study mass growth and dust formation/destruction mechanisms.
\end{abstract}



\section{Introduction} 
\label{sec:intro}

Over the last two decades, major advances have occurred  in assembling 
large galaxy samples from multiwavelength imaging surveys. Now the properties of 
galaxies are being studied with unprecedented accuracy in statistical representative samples 
of both local and distant galaxies. This has led to tremendous progress in improving 
our observational understanding on the evolution
of the galaxy stellar mass function \citep[GSMF, for recent discussions and compilations of observations up 
to high redshifts, see][]{Conselice+2016,Rodriguez-Puebla+2017}, as well as the evolution
of the star formation rate, SFR, as a function of stellar mass \ms\ \citep[see e.g.,][]{Noeske+2007,Elbaz+2007,Daddi+2007,Speagle+2014,Rodriguez-Puebla+2017}
and the cosmic star formation history \citep{Madau_Dickinson2014,Novak+2017,Driver+2018,Liu+2018}. 

With the above observational advances, the relationship between  
the $\sfr$ and \ms\ has received intense scrutiny  
by many authors over the last years. In particular, there is a class of  
star-forming galaxies, SFGs, that are known to obey a tight \sfr--\ms\ relation, 
exhibiting $\sim 0.3$ dex of scatter and establishing what in the literature has been 
coined as the ``Star-forming Main Sequence" 
\citep[MS,][for more recent studies and reviews see \citealp{Speagle+2014,Whitaker+2014,Santini+2017}]{Brinchmann+2004,Noeske+2007,Salim+2007}. 
The tightness of this relation and the abundance of galaxies on the MS, reflects that most galaxies 
spend a considerable time building their stars in an approximately self-regulated way {\it within} 
this sequence \citep[see e.g.,][]{Bouche+2010,Romeel+2012}, hence its importance for galaxy formation. 
Starburst galaxies are another class of SFGs with elevated amounts of star formation, typically 
associated with galaxy mergers, that are outliers of the MS but represent only a few 
percent ($\sim10-15\%$) of 
the total population \citep{Rodighiero+2011,Sargent+2012}. Quenched galaxies are another class of 
galaxies that are off the MS, in this case due their low levels of star formation 
\citep[e.g.,][]{Wuyts+2007,Williams+2009};  their fraction has increased rapidly
since $z\lesssim2$ \citep{Muzzin+2013}. In this paper we focus on MS
SFGs. 

Considering these now well-established findings, there are yet some open questions 
with respect to the distribution of the SFRs of SFGs. Does the \sfr--\ms\ relationship 
extend to low-mass galaxies and dwarfs, $\ms\lesssim10^{9}\msun$, with the same {\it slope}
as intermediate- and high-mass galaxies? 
What is the contribution from dust-obscured star formation as a function of stellar mass and redshift? 

This paper addresses the above questions by using the redshift evolution
of the far-ultraviolet FUV ($1500 \AA$) and the total infrared, IR ($8-1000\mu$m) rest-frame luminosity functions, LFs,
combined with the evolution of the GSMF of SFGs. 
The above questions 
have been studied separately by some authors in the past;
for example, \citet{Kurczynski+2016,McGaugh+2017,Iyer+2018} \citet{Davies+2019} studied the \sfr-\ms\ relation at low
masses while \citet{Pannella+2009} and \citet{Whitaker+2017} studied the contributions from dust-obscured star formation.
Normally, the approach employed by these and other authors is to determine those properties galaxy-by-galaxy  
from galaxy samples. However, the galaxy-by-galaxy approach fails to check consistency 
with other important statistical quantities 
that provide valuable constraints, such as, the GSMF and the FUV and IR LFs. Thus, it is key that robust determinations of the
total SFRs and the portions associated to the unobscured, \sfruv, and obscured, \sfrlir, regimes must be consistent
with the GSMF and the FUV and IR LFs. Specially,
it is important and timely to address the above two questions {\it at the same time} as well as their possible physical 
correlations and implications for galaxy formation models in the light of the joint evolution of the GSMF and the FUV and IR LFs.
Moreover, we note that the joint evolution of the GSMF and the FUV and IR rest-frame 
luminosity functions has not yet received attention in the light of the above questions. 
To our knowledge, there is only one study that exploits the joint evolution of the GSMF and the FUV and IR LFs, \citet{Bernhard+2014}. In that paper, the authors developed a phenomenological approach to estimate
the FUV and IR LFs based on the observed GSMF, the \sfr--\ms\ relationship and the dust attenuation, IRX.
While here we revolve around a similar idea our goal is different.\footnote{There are, however, other notable studies that have combined some of them 
but not all of them at the same time. For example, \citet{Bethermin+2012} used the IR LF and 
the GSMF while \citet{Burgarella+2013} used the FUV and IR. More recently \citet{Tacchella+2018}
used the FUV to study the distribution of SFRs.}


Our main goal for this paper is to exploit the {\it power} of using  
the joint evolution of the GSMF and the FUV and IR LFs 
as an alternative way to derive and study the \sfr--\ms\ relationship, and its decomposition into the unobscured and 
obscured contributions. 
Thus our approach exploits data not previously used to {\it derive} the \sfr--\ms\ relationship for MS SFGs. 
This is independent to the galaxy-by-galaxy approach which is based on galaxy 
surveys that could be subject to biases and incompleteness.
One of the main findings of this paper is that our results are in agreement with the galaxy-by-galaxy approach,
providing the kind of end-to-end test needed to verify the consistency between the SFRs,
including their unobscured and obscured contributions, with the GSMF and the FUV and IR LFs.
We also find that the joint evolution of the GSMF and the FUV and IR LFs offer a powerful tool that 
contributes to the study of the emergence  
of the galaxy mass distributions, stellar mass growth and dust formation/destruction, 
and ultimately to better constrain galaxy formation models. 

This paper is divided into four sections. The analytical method we employed to link \ms\ to $\sfr$ is 
introduced in Section \ref{sec:abundance_matching}. Section \ref{sec:results} presents our results and
compares to a set of independent observations finding a
good agreement between them. Our approach 
represents a new powerful and novel tool to study  at the same time the 
redshift evolution of MS SFGs
and the dependence of the fraction of dust-obscured star formation with stellar mass. 
Finally, in Section \ref{sec:discussion} we summarize and
discuss our results. 
 
In this paper we adopt the following cosmological parameter values: 
$\Omega_{\Lambda}  = 0.7$,  $\Omega_{\rm m}  = 0.3$,
 and $h = 0.7$. All stellar masses are normalized 
 to a \citet{Chabrier2003} Initial Mass Function, IMF.

\section{Modeling the FUV+IR SFR distributions} 
\label{sec:abundance_matching}

Here we present a mathematical approach to derive the 
redshift evolution of the conditional distribution of SFR given \ms\ for MS galaxies and all its moments. 
In particular, our main aim is the derivation of the redshift evolution of the mean \sfr--\ms\ relationship 
for MS galaxies. 
Normally, the evolution of the MS is inferred by combining galaxy samples at different redshifts for which the SFRs are estimated
for every galaxy  \citep[e.g.,][]{Whitaker+2014,Salim+2016}.
At this point, it should be said that  \textit{(i)} 
galaxy samples are subject to (their own) selection effects, 
and \textit{(ii)} the SFRs reported by different authors are based on different tracers and methods. 
As a result, the derived  SFR--\ms\ relation and its evolution will be biased. On the other hand, the comparison of
the SFR--\ms\ relation and its evolution estimated by different ways is not trivial, 
though some authors have attempted to homogenize different determinations
\citep[see e.g.,][]{Speagle+2014,Rodriguez-Puebla+2017}. 
Thus, to infer robust relations, specially for constraining accurately 
the shape of the MS in a {\it broad mass} and  {\it redshift} ranges, 
the optimal is to use a homogeneous method to calculate 
the SFRs from galaxy samples for which the volume completeness is well defined.  
This is the approach used here.

As mentioned earlier, the novelty of our approach is that it combines 
observational FUV and IR rest-frame LFs 
at different redshifts with the 
GSMFs of SFGs to derive the mean \sfr--\ms\ relationship and its scatter
over a   {\it broad mass} and {\it redshift} ranges. 
 The advantage of this approach is that it uses a uniform way to determine the SFRs at all redshifts and the sample completeness above given limits is well controlled.


Following, we describe how we constrain the conditional distribution of SFR given \ms\
when combining the FUV and IR LFs and the GSMF of SFGs at different redshifts.
Use of conditional distributions allows us to compute any moment, in particular 
the first moment, that is, the mean \sfr--\ms\ relationship. Notice that 
our method naturally separates the SFRs into their unobscured and dust-obscured components,
allowing study of the dust-obscured fraction at the same time.
 
\subsection{The FUV and IR conditional SFR distributions} 

Motivated by observations, we assume that the MS is a  non decreasing relation between \ms\ and \sfr.
The {\it key assumption} in our approach is that the FUV and IR SFR 
conditional probability distributions as a function of \ms\ convolved with  
the observed GSMF of SFGs gives
the FUV and IR rest-frame LFs. 
In other words, we are {\it assuming} that the FUV and IR LFs emerged from the MS galaxy population.
This is actually inaccurate in the case of the IR LF as starburst galaxies also contribute 
to the LF \citep[see e.g.,][]{Sargent+2012,Gruppioni+2013}. Fortunately,  
their contribution becomes relatively important only at the high luminosity end, and this can be estimated. 
Following the results presented in \citet{Sargent+2012}, we assume that the IR-based SFR distribution is composed
of the contributions of both MS and starburst galaxies, this is known in the literature as the 
2 star formation modes formalism. Next, we will discuss the 
conditional SFR distributions given \ms\ and their moments. While all the quantities that we discuss below
depend on redshift, we do not show this dependence for simplicity.

We begin by introducing the mean $\langle \log \sfr_{\rm FUV} (\ms) \rangle$ and $\langle \log \sfr_{\rm IR} (\ms) \rangle$ 
relations and our proposed parametric functions. 
To our knowledge there are not parametric functions previously reported in the literature for these relations.
We propose our parametric functions by noticing that the generic shapes of the 
\sfruv--\ms\ and  \sfrlir--\ms\ relations are governed by the shape of their corresponding FUV and IR LFs.
For the mean  \sfruv--\ms\  relation,  we assume a double power law (defined by four parameters: $ \mu_{{\rm FUV},0}$, $\alpha$, $\beta$ and $M'_{\rm FUV}$):
	\begin{equation}
		\begin{split}
			\langle\log\sfruv(\ms)\rangle = \mu_{{\rm FUV},0} \\
			+\log\left(\frac{2}{\left(\frac{M_{\ast}}{M'_{\rm FUV}}\right)^\alpha + \left(\frac{M_{\ast}}{M'_{\rm FUV}}\right)^\beta}\right),
		\end{split}
		\label{ec:sfruv_ms}
	\end{equation}
while for the mean \sfrlir--\ms\ we assume a Gompertz-like function that dominates at the low-mass end and 
a power-law that dominates at the high mass end (defined by three parameters: 
$ \mu_{{\rm IR},0}$, $\gamma$ and $M'_{\rm IR}$):
	\begin{equation}
		\begin{split}
			\langle\log\sfrlir(\ms)\rangle = \mu_{{\rm IR},0}  \\
			+ \log\left( \frac{ \left(M_{\ast}/M'_{\rm IR}\right)^\gamma }  {\exp\left( -e^{-\log(M_{\ast}/M'_{\rm IR})} \right)} \right).
		\end{split}
		\label{ec:sfrir_ms}
	\end{equation}
As mentioned above, all the parameters from the above equations depend on redshift.
In Section \ref{secc:constrains} we discuss the functions that we use for the dependences with redshift and
their corresponding best fitting parameters to the observational data through our approach.

To describe the 
conditional SFR distributions, we introduce the conditional log-normal distribution 
$\mathcal{N}(y|x)$ given by:
	\begin{equation}
		\mathcal{N}(y|x)d\log y = \frac{d\log y}{\sqrt{2\pi\sigma_y^2}}
		 \exp\left[-\frac{(\log y - \mu (x))^2}{2\sigma_y^2}\right],
		  \label{eq:lognormal}
	\end{equation}
where the mean is $ \mu (x) = \langle \log y (x)\rangle$, and $\sigma_y$ is the logarithmic width of the
distribution, i.e., its scatter. Since we will work in a space in which 
the logarithm of  $y$ is log-normally distributed, for simplicity, we will refer sometimes to $ \langle \log y (x)\rangle$
as the mean $y-x$ relation. In some cases authors report $\langle y\rangle$; 
for a log-normal distribution the offset between $ \langle \log y\rangle$ and $\langle y\rangle$  
 is given by \citep[see e.g.,][]{Rodriguez-Puebla+2017}: 
	 \begin{equation}
		 \log \langle  y\rangle = \langle\log  y\rangle +  \frac{\sigma^2}{2}\ln 10.
		 \label{ec:mean_diff}
	\end{equation}

We define the conditional probability distribution of SFRs given a stellar mass as 
$\mathcal{P}_{{\rm SFR},X}({\rm SFR}_X|\ms)$, where $X$ refers to the SFRs computed based on
the FUV or IR rest-frame luminosities. 
In the case of \sfruv, we assume that 
$\mathcal{P}_{{\rm SFR},\rm FUV} = \mathcal{N}(\mu_{\rm FUV}, \sigma_{\rm FUV})$
with a scatter $\sigma_{\rm FUV}$ independent of \ms\ and mean 
$\mu_{\rm FUV} = \langle\log\sfruv(\ms)\rangle$, given by Equation (\ref{ec:sfruv_ms}). As for the distribution of \sfrlir, 
as mentioned above, it is composed of the contributions of MS and starburst galaxies.
Therefore,
	\begin{equation*}
		\mathcal{P}_{{\rm SFR},\rm IR} = (1 - \mathcal{A}_{SB})\times \mathcal{P}_{{\rm SFR},\rm IR-MS} +   \mathcal{A}_{SB}\times \mathcal{P}_{{\rm SFR},\rm IR-SB},
	\end{equation*} 
where $\mathcal{A}_{SB}$ is the fraction of starburst galaxies at fixed stellar mass, and the conditional distributions 
$\mathcal{P}_{{\rm SFR},\rm IR-MS}$ 
and $\mathcal{P}_{{\rm SFR},\rm IR-SB}$ are log-normally distributed (Eq. \ref{eq:lognormal}). Following \citet{Sargent+2012}, 
we assume that $\sigma_{\rm IR} (= \sigma_{\rm IR-SB} = \sigma_{\rm IR-MS})$ 
and $\mathcal{A}_{SB}$ are both independent of mass and that $\mu_{\rm IR-SB} = \mu_{\rm IR-MS} + \Delta \mu_0$,
where $\mu_{\rm IR-MS} = \langle\log {\rm SFR_{\rm IR-MS}}(\ms)\rangle$ 
is the mean relation for the MS galaxies (give by Equation (\ref{ec:sfrir_ms}), 
hereafter we will refer to it just as $\langle\log {\rm SFR_{\rm IR}}\rangle$).
Based on the recent results from \citet{Schreiber+2015} we use $\mathcal{A}_{SB} = 0.033$ and 
$\Delta \mu_0 = 0.79$ dex.\footnote{Note that this value slightly differs from the one reported in  \citet{Schreiber+2015}. 
The reason for this is that we are taking into account that the distribution between the ratio of the sSFRs and the mean
\ssfr\ from the MS
($R_{\rm SB}$, the starburstiness parameter defined by the authors)
is not centred on $R_{\rm SB} = 1$, 
see their best fit models to their equation 10 and figure 19.}

As for the scatters $\sigma_{\rm FUV}$ and $\sigma_{\rm IR}$, first note that  
observations based on mass-complete samples have shown that the scatter around the 
 \sfr--\ms\ relationship is of the order of $0.3$ dex \citep[][]{Speagle+2014,Schreiber+2015,Cano-Diaz+2016,Fang+2018}. 
 Nonetheless, recent results have shown that the scatter 
could increase up to $\sim0.4$ dex at the high mass end \citep{Ilbert+2015,Popesso+2019,Popesso+2019a}. 
While the discussion of the real trends and values of the MS scatter is beyond the scope of this paper, 
 here we assume a constant scatter of $\sigma_{\rm FUV} = 0.3$ dex for the FUV and $\sigma_{\rm IR} = 0.4$ 
 for the IR and, for simplicity, zero covariance between the two distributions. 
The above will lead to a scatter around the inferred mean SFR--\ms\ relations that has
some dependence on \ms\ (see Equation \ref{eq:scatter_logsfr} described below), in the direction 
that for low-mass galaxies,  $\sigma\sim0.3$ dex while for massive galaxies, $\sigma\sim0.4$ dex. 

\subsection{The FUV and IR LFs} 

In this section we describe how the FUV and IR LFs are derived by convolving  
the GSMF of SFGs and the conditional FUV and IR SFR probability distributions.
For the above, we begin by using the \citet{Kennicutt1998} conversion factors,  $\mathcal{K}_X$, to transform SFRs into 
luminosities, that is,  $ L_X = \mathcal{K}_X/ {\rm SFR}_X$.
These conversion factors allow us to 
pass from the SFR conditional probability distributions, $\mathcal{P}_{{\rm SFR},X}({\rm SFR}_X|\ms)$, 
to the respective distributions of luminosities: $\mathcal{P}_{l,X}(L_X|\ms)$.
Then the LF of the FUV and IR light is related to the GSMF of SFGs, $\phi_{\ast,{\rm SF}}$, by the following convolution:
	\begin{equation}
		\phi_{X}(L_{X})  =   \int \mathcal{P}_{l,X}(L_X|\ms)\phi_{\ast,{\rm SF}}(\ms) d\log \ms.
		\label{ec:LFs_convolution}
	\end{equation}
For the GSMF of SFGs we use the best fitting model to the compiled GSMF data from \citet{Rodriguez-Puebla+2017}
and their reported evolution of the fraction of SFGs as a function of 
\ms\ (for details see Appendix \ref{secc:gsmf} ). 
We note that the above integral is not computed from $\ms = 0$ but 
from $\ms = 10^{6}\msun$, which is yet $\sim1-1.5$ dex below the observational limit of our GSMFs.

The observational LFs  are actually reported over redshift intervals, $z_i< z< z_f$. We take into
account the above by using the following equation:
	\begin{equation}
		\phi_{{\rm obs},X}(L_{X};z_i,z_f)  =\frac{\int_{z_i}^{z_f} \phi_{X}(L_{X},z) dV}{V(z_f) - V(z_i)},
		\label{ec:phi_conv}
	\end{equation}
where $V$ is the comoving volume. 

The original  \citet{Kennicutt1998} conversion factors were obtained on the assumption of a 
\citet{Salpeter1955} IMF.\footnote{Using the conversion factors is
a very popular approach to empirically relate SFRs to luminosities. While there is some justification in using this 
method, it may nevertheless yield values that are different from the real ones. 
For example, the conversion factor of the SFR diagnostic based on the FUV light depends on 
the metal-enrichment and the star formation history, SFH, of a galaxy. This could 
lead to differences up to $\sim0.5$ dex in the 
conversion factor when exploring different combinations of these dependencies \citep[see figure 3 from][see also \citealp{Tacchella+2018}]{Madau_Dickinson2014}. 
While the above introduces  an uncertainty that could be taken into account in our analysis, 
we decided to use the most popular calibration factors from the literature because this facilitates the comparison with previous 
works. Therefore, we do not consider more complex conversion factors that depend on metallicity and SFHs.}  
According to \citet{Madau_Dickinson2014}, we multiply the original 
conversion factors by 0.63 to convert them into a \citet{Chabrier2003} IMF:
the FUV conversion factor changes to 
$\kfuv = 1.7 \times 10^{-10}\msun$ yr$^{-1}L_{\odot}^{-1}$, while the IR factor changes to $\kir = 1.09 
\times 10^{-10}\msun$ yr$^{-1}L_{\odot}^{-1}$. Notice that the FUV light 
refers to the {\it observed luminosity}, which can 
be obscured in the presence of dust. 
In the following, we assume that the fraction of FUV light, emitted by young stars,
that is absorbed by dust is directly proportional to the fraction that is reemitted in the IR 
(wavelength range $8-1000\mu$m), as is often assumed in the literature \citep{daCunha+2008}.

\begin{figure*} 
	\vspace*{-0pt}
	\hspace*{0pt}
	\includegraphics[height=5.3in,width=7.3in]{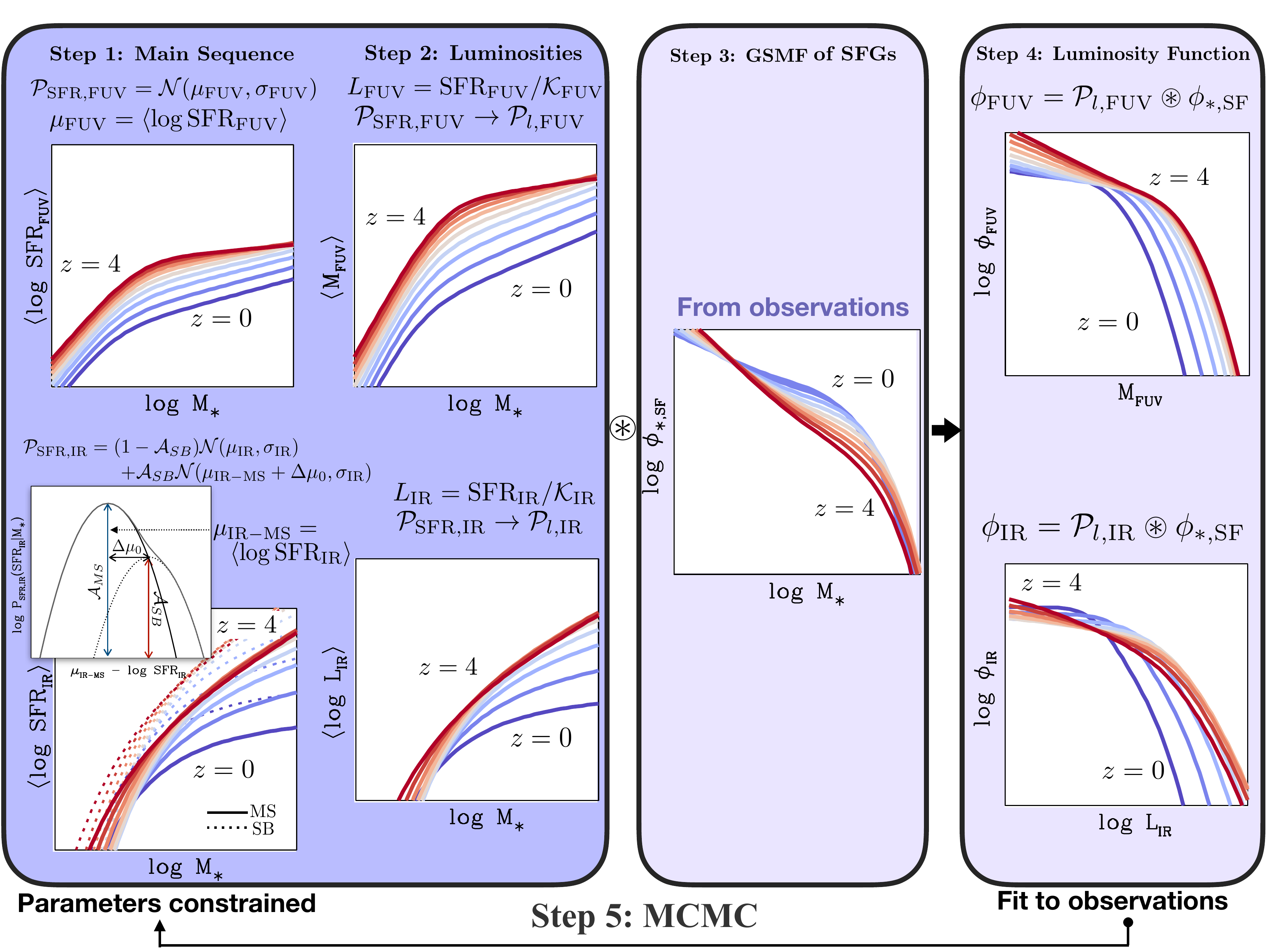}
		\caption{Scheme with the steps to constrain the means of SFR given \ms\ at different
		redhsifts from
		fitting the observational FUV and IR LFs and knowing the GSMFs of SFGs; see text for details.} 
	\label{fig:cartoon} 
\end{figure*}

\subsection{Constraining the FUV+IR SFR distributions}
\label{secc:constrains}

Under the assumption of a monotonic increasing relation between $L_X$ ($X$=FUV or IR) and \ms, Eq. (\ref{ec:LFs_convolution}) shows how this relation, or more generally, the respective luminosity conditional probability distribution,
maps the GSMF of SFGs into the FUV or IR LF.  These FUV and IR luminosity distributions are tracers of the 
respective SFR distributions.
Our next aim is to constrain self-consistently these SFR distributions (FUV, IR, and the total) from observations.
Figure \ref{fig:cartoon} summarizes the steps described in the previous subsections and here.
The steps are as follows:

	\begin{description}
		\item[Step 1]	Define the FUV SFR conditional probability distribution $\mathcal{P}_{{\rm SFR},\rm FUV}$
					as a log-normal function.  Use the two-star-formation-mode from \citet{Sargent+2012} 
					to separate the contribution from MS and starburst galaxies. Define the IR SFR 
					 conditional probability distribution as  
					 $\mathcal{P}_{{\rm SFR},\rm IR} = (1-\mathcal{A}_{SB})\mathcal{P}_{{\rm SFR},\rm IR-SF}+
					 \mathcal{A}_{SB} \mathcal{P}_{{\rm SFR},\rm IR-SB}$. Both for $\mathcal{P}_{{\rm SFR},\rm IR-SF}$ and
					$\mathcal{P}_{{\rm SFR},\rm IR-SB}$ are assumed to be log-normal functions.
%
					Propose redshift-depedent parametric functions for the
					mean relations $\langle\log\sfruv(\ms)\rangle$ and  
					$\langle\log\sfrlir(\ms)\rangle$, Eqs. (\ref{ec:sfruv_ms}) and (\ref{ec:sfrir_ms}), respectively.

		\item[Step 2]	Convert SFRs into luminosities using the \citet{Kennicutt1998} 
					conversion factors. This allows us to 
					pass from the conditional distributions 
					$\mathcal{P}_{{\rm SFR},\rm FUV}$ and $\mathcal{P}_{{\rm SFR},\rm IR}$ 
					to the conditional distributions $\mathcal{P}_{l,\rm FUV}$ and  
					$\mathcal{P}_{l,\rm  IR}$, respectively.
					
		\item[Step 3]	Characterize the redshift evolution of the observational GSMF of SFGs, $\phi_{\ast,{\rm SF}}$. 
					Appendix \ref{secc:gsmf} presents our best fitting model to the compiled GSMFs and 
					the fraction of SFGs as a function of redshift from \citet{Rodriguez-Puebla+2017}.

		\item[Step 4]	Derive the FUV and IR LFs by convolving the GSMF of SFGs with their corresponding 
					conditional luminosity distributions (Eq. \ref{ec:LFs_convolution}):
					$\phi_{\rm FUV} = \mathcal{P}_{l,\rm FUV} \circledast \phi_{\ast,{\rm SF}}$ and 
					$\phi_{\rm IR} = \mathcal{P}_{l,\rm IR}  \circledast \phi_{\ast,{\rm SF}}$.
					
		\item[Step 5]	Adjust the redshift-dependent parameters 
		                         of the mean $\langle\log\sfruv(\ms)\rangle$ and  
					$\langle\log\sfrlir(\ms)\rangle$ relations. This is possible since steps 1-4 
					relate the mean relations to the observed FUV and IR LFs. 
					
	\end{description}

\begin{deluxetable}{lC}
	\tablecaption{FUV LFs}\label{tab:T1}
	\tablewidth{0pt}
	\tablehead{
	\colhead{Author} & \colhead{Redshift}
	}
	\startdata
	\citet{Cucciati+2012} & 0.05  < z  < 4.5\\  
	\citet{Arnouts+2005} & 0.055 < z < 3.5\\ 
	\citet{Driver+2012} & 0.013< z < 0.1\\
	\citet{Robotham_Driver2011} & 0.013 < z < 0.1\\  
	\citet{Oesch+2010} & 0.75  \lesssim z \lesssim 2.5\\  
	\citet{Alavi+2016} & 1< z < 3\\	
	\citet{Hathi+2010} & 1 < z < 3\\ 
	\citet{Mehta+2017} & 1.4 < z < 3.6\\ 	
	\citet{Reddy+2009} & 1.9 < z < 3.4\\ 
	\citet{Alavi+2014} & z\sim2\\	
	\citet{Parsa+2016} & 2 < z < 4\\ 	
	\citet{vanderBurg+2010} & 3  < z < 5\\  
	\citet{Ono+2018} & 4 < z < 7\\  
	\citet{Bouwens+2007} & 4  \lesssim z  \lesssim 5\\ 
	\citet{Finkelstein+2015} & 4  \lesssim z  \lesssim 8\\  
	\citet{Bouwens+2015} & 4  \lesssim z  \lesssim 8\\ 
	\citet{Atek+2018} & z\sim6\\ 
	\citet{Livermore+2017} & 6 < z < 8\\ 
	\citet{Bhatawdekar+2019} & 6 < z < 9\\ 
	\citet{Bouwens+2011} & 7  \lesssim z  \lesssim 8\\  
	\citet{Oesch+2012} & z\sim 8\\  
	\enddata
\end{deluxetable}



Next, we describe the parametrization and the best fitting model to the redshift evolution of the
mean \sfruv--\ms\ and \sfrlir--\ms\ relations.

Each of the parameters described by Equations (\ref{ec:sfruv_ms}) and (\ref{ec:sfrir_ms})
depend on redshift. After some experimentation, we found that the redshift $z$ dependence of 
these parameters is well described by the following function composed of 3 free parameters:
	\begin{equation}
		\mathcal{Q} (q_1,q_2,q_3,z)= q_1 + q_2 \log(1+z) + q_3 z^2.
	\end{equation}
In this paper, we find the best  fitting parameters by using a Bayesian approach through a MCMC method applied jointly to the data 
following \citet{Rodriguez-Puebla+2013}. 

\begin{figure*} 
	\vspace*{-60pt}
	\hspace*{40pt}
	\includegraphics[height=7in,width=6in]{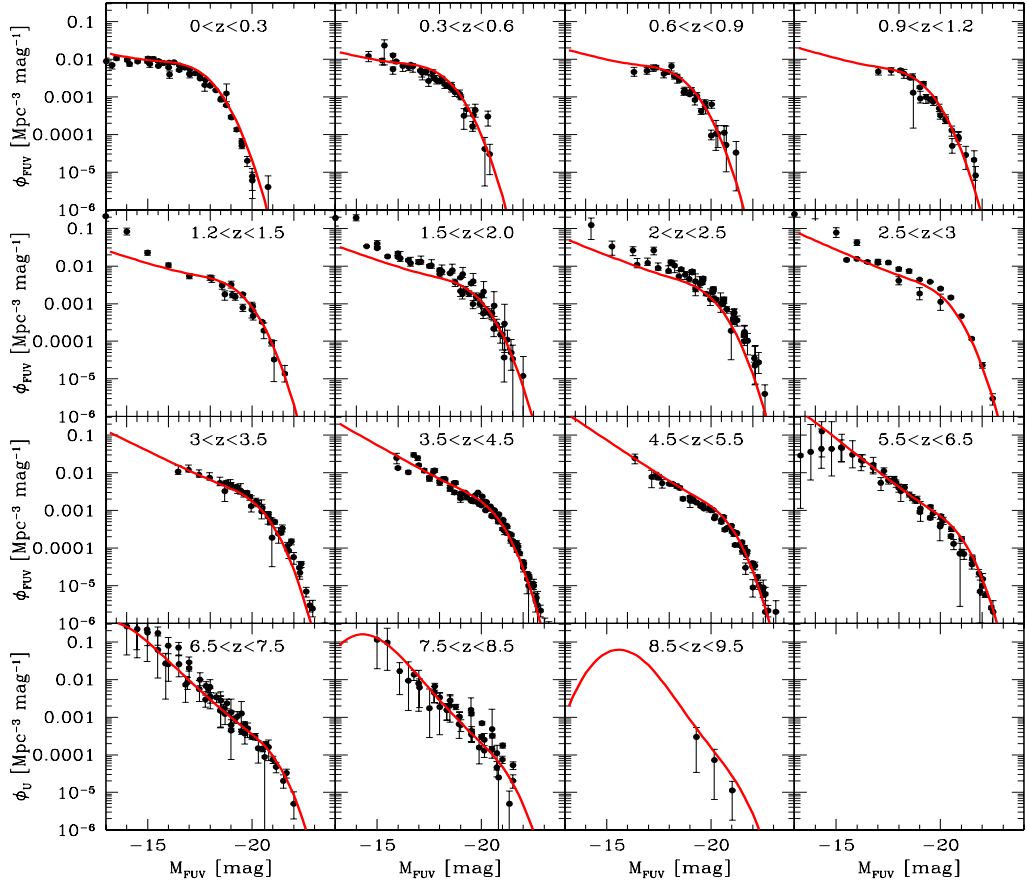}
	\vspace*{-100pt}
		\caption{FUV Luminosity Function. The red solid lines present the best fitting models to a compilation of 
		21 observational studies, see Table \ref{tab:T1}. The data has been homogenized to a same cosmology. 
		The turn over at the faint end at $z>7.5$ is due to the lower limit of \ms\ considered for the 
		integral in Equation (\ref{ec:LFs_convolution}).} 
	\label{fig:uv_lf} 
\end{figure*}

Figure \ref{fig:uv_lf} shows a compilation of measured rest-frame FUV LFs between $z=0$ and $z=9$ 
from 21 observational studies that are listed in Table \ref{tab:T1}. The FUV LFs were not corrected for dust attenuation but
homogenized to our adopted cosmology.
We use 
this compilation to find the best fit parameters to the evolution of the mean \sfruv--\ms\ relationship. 
The solid lines in Figure \ref{fig:uv_lf} show our best fitting model to the LF 
$\phi_{{\rm obs, FUV}}$. The best fit parameters of the redshfit-dependent mean \sfruv-\ms\ relationship are:
	\begin{equation}
		\begin{split}
		\mu_{{\rm FUV},0} (z) = \mathcal{Q}(-1.451\pm0.011,2.471\pm0.027,\\
		-0.001\pm0.0004,z),
		\end{split}
	\end{equation}
	\begin{equation}
		\begin{split}
		\log M_{\rm FUV} (z) = \mathcal{Q}(8.343\pm0.011,0.705\pm0.030,\\
		0.002\pm0.0007,z),
		\end{split}
	\end{equation}
	\begin{equation}
		\begin{split}
		\alpha(z)  = \mathcal{Q}(-0.543\pm0.009,1.978\times10^{-5}\pm0.002,\\
		0.010\pm0.0002,z),
		\end{split}
	\end{equation}
	\begin{equation}
		\begin{split}
		\beta(z)  = \mathcal{Q}(0.667\pm0.006,0.033\pm0.015,\\
		0.012\pm0.0004,z).
		\end{split}
	\end{equation}

\begin{figure*} 
	\vspace*{-240pt}
	\hspace*{35pt}
	\includegraphics[height=8.in,width=6in]{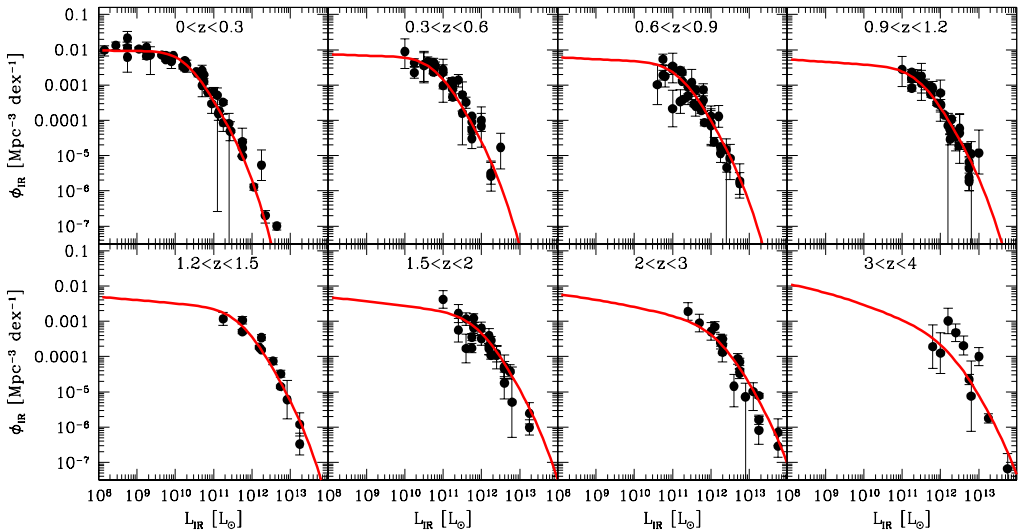}
	\vspace*{-120pt}
		\caption{IR Luminosity Function. Solid lines present the best fit from a compilation of 
		8 observational studies, see Table \ref{tab:T2}. Similarly to the FUV LFs, the data has been homogenized to a same cosmology } 
	\label{fig:ir_lf} 
\end{figure*}

The observed IR LFs from $z=0$ to $z=4.2$ are shown in Figure \ref{fig:ir_lf} from a compilation of 8 observational studies 
listed in Table \ref{tab:T2}. Similarly above the data was homogenized to our adopted cosmology.  
Using these data, as in the case of the FUV data, we constrain the  best fit parameters of 
the redshift-dependent mean \sfrlir--\ms\ relationship:
	\begin{equation}
		\begin{split}
		\mu_{{\rm IR},0} (z) = \mathcal{Q}(0.127\pm0.017,5\pm0.044,\\
		-0.168\pm0.028,z),
		\end{split}
	\end{equation}
	\begin{equation}
		\begin{split}
		\log M_{\rm IR} (z) = \mathcal{Q}(9.850\pm0.031,1.876\pm0.087,\\
		-0.107\pm0.030,z),
		\end{split}
	\end{equation}
	\begin{equation}
		\gamma(z)  = \mathcal{Q}(0.051\pm0.011,0.983\pm0.066,0,z).
	\end{equation}
Our best fitting model is shown with the red solid lines in Figure \ref{fig:ir_lf}.  

\begin{deluxetable}{lC}
	\tablecaption{IR LFs}\label{tab:T2}
	\tablewidth{0pt}
	\tablehead{
	\colhead{Author} & \colhead{Redshift}
	}
	\startdata
	\citet{KilerciEser+2018} & 0< z < 0.3\\
	\citet{LeFloc'h+2005} & 0< z < 1\\
	\citet{Casey+2012} & 0< z < 1.6\\
	\citet{Rodighiero+2010b} & 0<z<2.5\\
	\citet{Lim+2020} & 0< z < 4\\
	\citet{Gruppioni+2013} & 0< z < 4.2\\
	\citet{Magnelli+2013} & 0.1< z < 2.3\\
	\citet{Magnelli+2011} & 1.3< z < 2.3\\
	\enddata
\end{deluxetable}

\subsection{The total SFR distribution: the mean SFR--\ms\ relation and the scatter}

Once the redshift-dependent parameters of the mean 
$\langle\log\sfruv(\ms)\rangle$ and  $\langle\log\sfrlir(\ms)\rangle$ relations are constrained, 
we now compute the total (FUV+IR) \sfr--\ms\ relations  and the scatter around it. 
Thus, the final step in our program is to characterize the total conditional SFR 
distribution at a fixed \ms\ for galaxies from the FUV- and IR-based conditional SFR distributions. 
Formally, the total SFR distribution is given by:
	\begin{equation}
		\begin{split}
		\mathcal{P}_{{\rm SFR}}(\sfr|\ms) = \int  \frac{ \mathcal{P}_{{\rm SFR},\rm FUV}({\rm SFR}-{\rm SFR_{IR}}|\ms)}{1 - {\rm SFR_{IR}}/{\rm SFR} }  \times  \\
		 \mathcal{P}_{{\rm SFR},\rm IR-MS}(\sfrlir|\ms) d\log \sfrlir .\
		\label{ec:P_sfrs}		 
		\end{split}
	\end{equation}
Note that for the IR SFR distribution we use only the component of MS galaxies. 

A few words on the above distribution are worth of mentioning at this point. 
While the sum of normally distributed random variables 
is also normally distributed, this is not the case for log-normal distributed random variables. 
To our knowledge there is not an analytical solution for log-normal distributions, so this is 
computed numerically via Equation (\ref{ec:P_sfrs}).
From the above equation we can derive the total mean (hereafter we will refer sometimes to it as simply 
the mean) SFR--\ms\ relation:
 	\begin{equation}
		\langle \log \sfr(\ms) \rangle = \int  \log \sfr\: \mathcal{P}_{{\rm SFR}}(\sfr|\ms) d\log\sfr.
		\label{eq:mean_logsfr}
	\end{equation}
The scatter around the mean is
 	\begin{equation}
		\begin{split}
			\sigma_{\rm SFR}^2 (\ms)= \int \left( \log \sfr - \langle \log \sfr(\ms) \rangle \right)^2 \times \\ 
			\mathcal{P}_{{\rm SFR}}(\sfr|\ms) d\log\sfr .\
		\label{eq:scatter_logsfr}			
		\end{split}
	\end{equation}
We will also report the mean $\langle \log \ssfr(\ms) \rangle$ 
relation with $\ssfr = \sfr/\ms$ and $\sigma_{\rm sSFR} =\sigma_{\rm SFR}$. 
Since we use separately the observations of the IR and FUV LFs, a straightforward outcome 
from our approach is the determination of the separate contributions from 
FUV and IR to the total SFRs. This will be important for deriving 
 the dust-obscured fraction  as a function of mass  and redshift.

\section{Results} 
\label{sec:results}

\begin{deluxetable*}{lcCCCC}
	\tablecaption{Star Formation Rates}\label{tab:T3}
	\tablewidth{0pt}
	\tablehead{
	\colhead{Author} & \colhead{SFR indicator} & \colhead{$\Delta_{\rm SFR} [{\rm dex}]$} & \colhead{$\Delta_{\rm M_{\ast}} [{\rm dex}]$} & \colhead{$\sigma_{\rm SFR} [{\rm dex}]$} & \colhead{Redshift}
	}
	\startdata
	\citet{Popesso+2019,Popesso+2019a} & UV+IR & 0 & 0 & 0.25-0.4 & 0<z < 2.5\\
	\citet{Lee+2015} & UV+IR & -0.15 & 0 & 0.36 & 0.2< z < 1.3\\
	\citet{Ilbert+2015} & UV+IR & 0 & 0 & 0.22-0.48 & 0.2< z < 1.4\\
	\citet{Whitaker+2012} & UV+IR & -0.15 & -0.03 & 0.34 & 0.5<z < 2.5\\
	\citet{Whitaker+2014} & UV+IR & -0.13 & 0 & 0.34^{a,b} & 0.5< z < 2.5\\
	\citet{Tomczak+2016} & UV+IR & -0.13 & 0 & 0.34^{a,b} & 0.5< z < 4\\
	\citet{Schreiber+2015} & UV+IR & -0.31 & -0.2 & 0.31 & z < 4\\
	\citet{Reddy+2012} & UV+IR \& SED &  -0.2 & -0.2 & 0.37 & 1.4<z < 3.7\\
	\citet{Tasca+2015} & UV &  0 & 0 & \sim0.54^{c} & 0.4\lesssim z\lesssim 4.8\\
	\citet{Bouwens+2012} & UV &  -0.3 & -0.2 & 0.3^{a,d} & z\sim4\\
	\citet{Gonzalez+2011} & UV &  -0.2 & -0.2 & 0^{a} & 4\lesssim z\lesssim 6\\
	\citet{Salmon+2015} & UV &  -0.2 & -0.2 & 0.25-0.42 & 4\lesssim z\lesssim 6\\
	\citet{Duncan+2014} & UV & 0 & 0 & 0.17-0.36 &  4\lesssim z\lesssim 6\\
	\citet{Khusanova+2020} & UV & 0 & 0 & 0^{a} & 5.5<z <6.6\\
	\citet{Karim+2011} & 1.4 GHz & +0.18$^{e}$ & -0.02 & 0^{a} & 0.2< z < 0.4\\
	\citet{Karim+2011} & 1.4 GHz & +0.15$^{e}$ & -0.02 & 0^{a} & 0.4< z < 0.6\\
	\citet{Karim+2011} & 1.4 GHz & +0.11$^{e}$ & -0.02 & 0^{a} & 0.6< z < 0.8\\
	\citet{Karim+2011} & 1.4 GHz & +0.07$^{e}$ & -0.02 & 0^{a} & 0.8< z < 1\\
	\citet{Karim+2011} & 1.4 GHz & +0.03$^{e}$ & -0.02& 0^{a} & 1< z < 1.2\\
	\citet{Karim+2011} & 1.4 GHz & -0.01$^{e}$ & -0.02 & 0^{a} & 1.2< z < 1.6\\
	\citet{Karim+2011} & 1.4 GHz & -0.07$^{e}$ & -0.02 & 0^{a} & 1.6< z < 2.0\\
	\citet{Karim+2011} & 1.4 GHz & -0.12$^{e}$ & -0.02 & 0^{a} & 2.0< z < 2.5\\
	\citet{Leslie+2020} & 3 GHz &  -0.1 & 0 & 0^{a,d} & 0.3< z < 6\\
	\enddata
	\tablecomments{$\Delta_{\rm SFR}$ and $\Delta_{\rm M_{\ast}}$ are the logarithmic correction offsets. 
	Correction to a same IMF
	and when authors report  $\langle \sfr\rangle$ instead of $\langle\log \sfr\rangle$ were applied.\\ 
	$^{a}$ No measure of the scatter available.\\
	$^{b}$ We assumed $\sigma_{\rm SFR} = 0.34$ as in \citet{Whitaker+2012}.\\
	$^{c}$ The authors reported that error bars were computed as $\sigma/\sqrt{N}$ where $\sigma$ is the standard deviation 
	around the sSFR distribution and $N$ the number of galaxies. Based on their values reported for the error bars we find that on
	average $\sigma\sim0.54$ dex. \\
	$^{d}$ We assumed $\sigma_{\rm SFR} = 0.3$.\\
	$^{e}$ Corrections taken from \citet{Speagle+2014}\\
	}
\end{deluxetable*}

Before discussing our results,
we note that the observed IR LF is the main source of uncertainty in 
the conclusions that will be presented below. This is due to its restricted 
redshift, $z\lesssim4$, and luminosity range, see Figure \ref{fig:ir_lf}. 
This is not the case for the FUV LF and the  GSMF,
which are available up to $z\sim9$ and cover several orders of magnitude 
from low- to high-luminosity/mass galaxies, see Figures 
\ref{fig:uv_lf} and also Figure \ref{fig:gsmf} from Appendix \ref{secc:gsmf}. 
In order to reduce the risk of over-interpreting our results, 
in the figures below we will indicate explicitly the regimes over which our IR LFs are valid
(the exception will be Figure \ref{fig:fraction_obsc}, but this is discussed in the text).

\begin{figure*}
	\vspace*{-320pt}
	\hspace*{-20pt}
	\includegraphics[height=10.7in,width=7.2in]{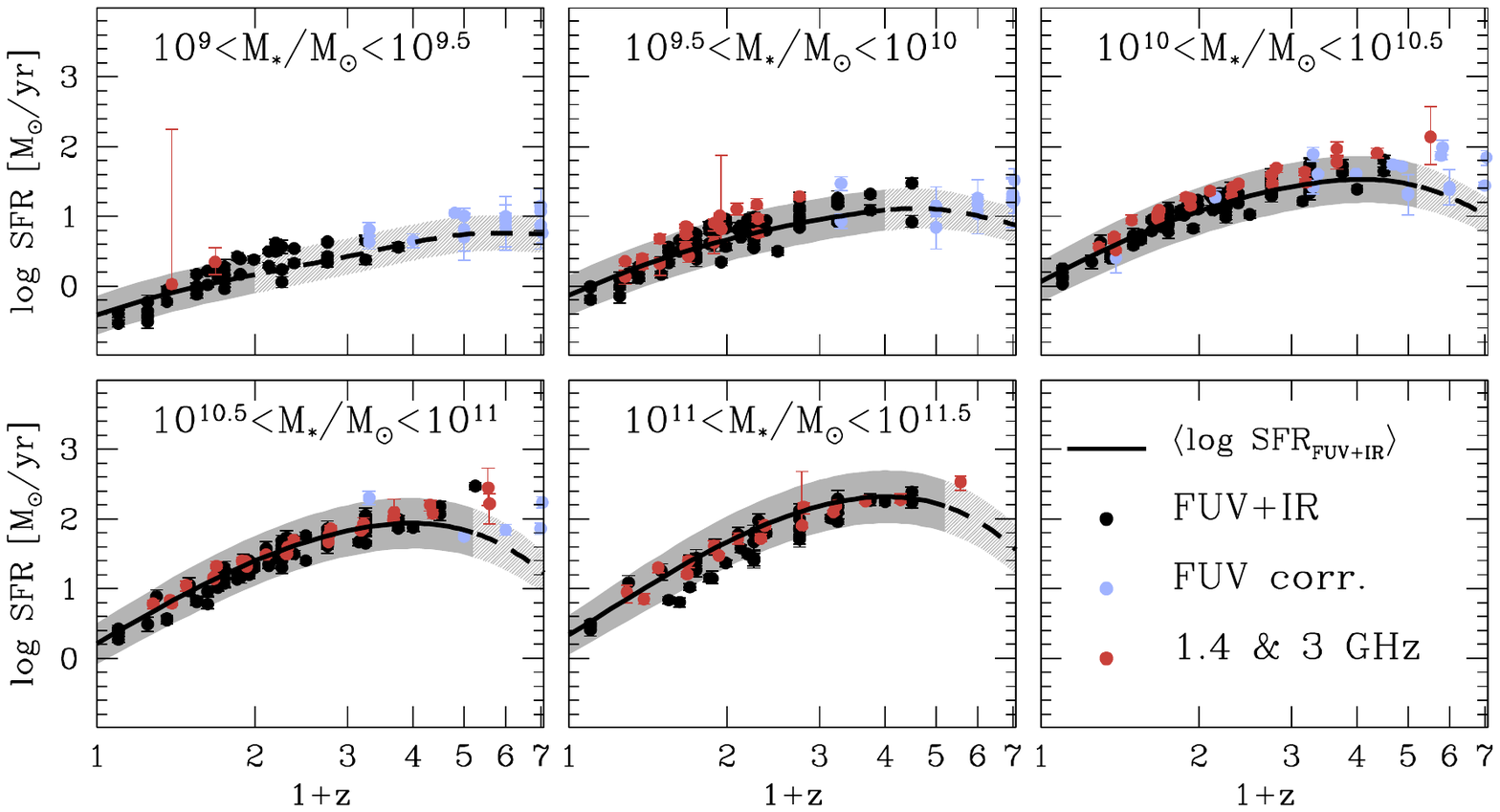}
	\vspace*{-170pt}
		\caption{SFRs as a function of redshift for five stellar mass bins.
		The solid lines show our results when the best fitting models of both 
		the FUV and IR LFs are available. The dashed lines show the mass regimes
		at the redshifts where the IR LFs have been extrapolated but the FUV LFs are available. 
		The filled circles with error bars 
		show the compilation reported from Table \ref{tab:T3} for SFGs from
		galaxy surveys. Black circles show the results based on FUV+IR, while
		light blue and red circles show the results based on FUV dust-corrected and
		radio data. Note that this figure shows that our estimates of total SFRs inferred 
		from fitting the FUV and IR LFs are in good agreement with direct measurements of the total SFR.
 	}
	\label{fig:ssfr_vs_z}
\end{figure*}

In Figure \ref{fig:ssfr_vs_z} we present our resulting  
$\langle \log \sfr\rangle$, see Eq. (\ref{eq:mean_logsfr}), as a function of redshift for five stellar mass bins. 
The solid lines indicate the stellar mass 
ranges when the best fitting models of both the FUV and IR LFs are constrained by the data. 
The dashed lines indicate the range over which the IR LFs have been extrapolated but there are observational data
to constrain the best fit models of the FUV LF. The shaded area shows the scatter around the mean relation
using Equation (\ref{eq:scatter_logsfr}). Note that the above limitation in the data affects mostly low-mass galaxies, 
$\ms\lesssim10^{9}\msun$. 

\begin{figure}
	\vspace*{-200pt}
	\hspace*{-10pt}
	\includegraphics[height=7.5in,width=5.8in]{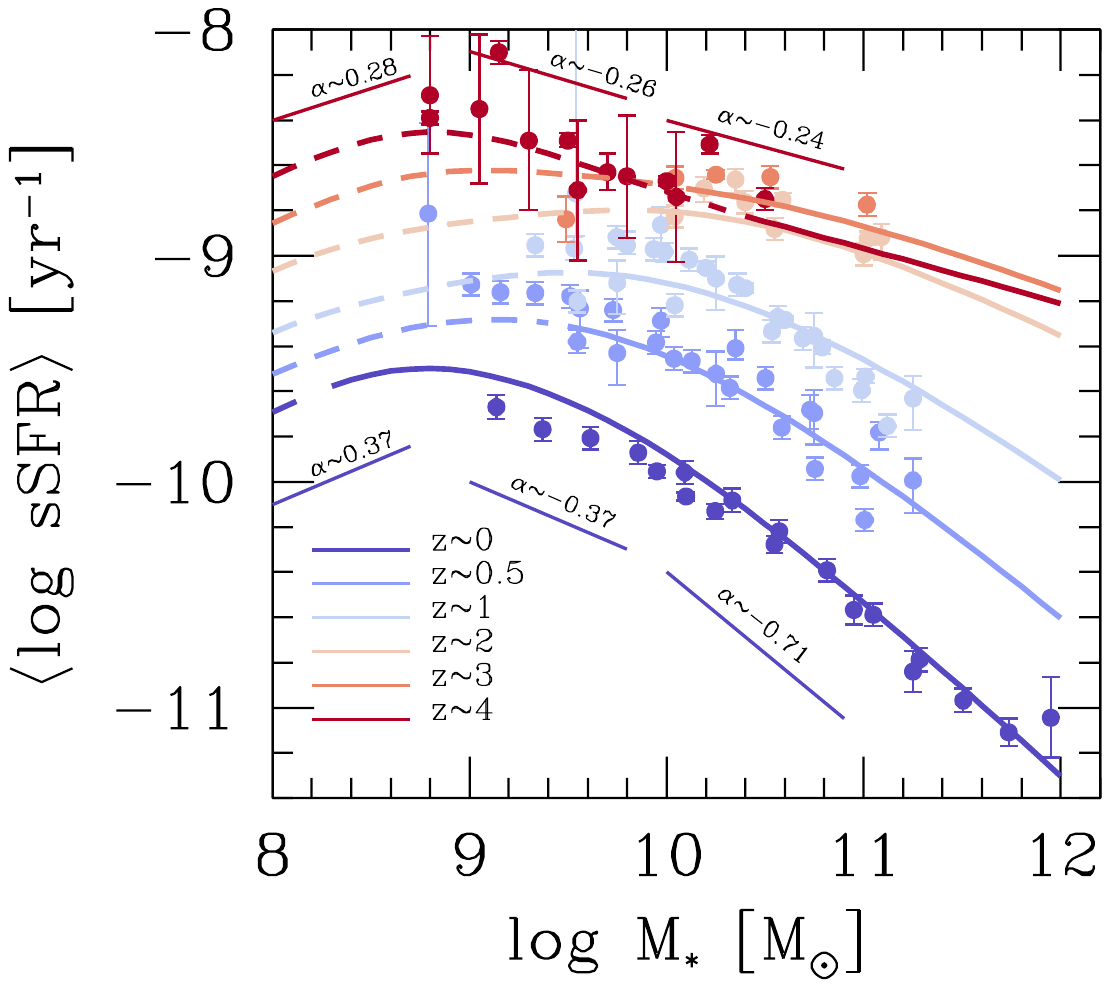}
	\vspace*{-120pt}
		\caption{The \ssfr-\ms\ relation for MS SFGs from $z\sim0$ to $z\sim4$.
		The solid lines show our results when the best fitting models of both 
		the FUV and IR LFs are available. The dashed lines show the mass regimes
		when the IR LFs have been extrapolated but the FUV LFs are available. 
		The filled circles with error bars 
		show the compilation from Table \ref{tab:T3} for MS SFGs. Our results are consistent 
		with the inferences from galaxy surveys. 
		The slope, $\alpha=d\langle\log\ssfr\rangle / d\log\ms$, 
		at $z\sim0$ for low masses, $\ms=10^{8}-10^{8.8}\msun$ is $\alpha\sim0.37$
		while for intermediate, $\ms=10^{9}-10^{10}\msun$, and high masses, $\ms=10^{10}-10^{12}\msun$, 
		we find respectively that $\alpha\sim-0.37$ and $\sim-0.71$.  
		At $z\sim4$, the slopes for intermediate and high masses the slopes 
		are respectively $\alpha\sim-0.26$ and $\alpha\sim-0.24$ while at low masses is $\alpha\sim0.28$.
 	}
	\label{fig:ssfr_vs_ms}
\end{figure}

\begin{figure}
	\vspace*{-50pt}
	\hspace*{-20pt}
	\includegraphics[height=7.1in,width=5.3in]{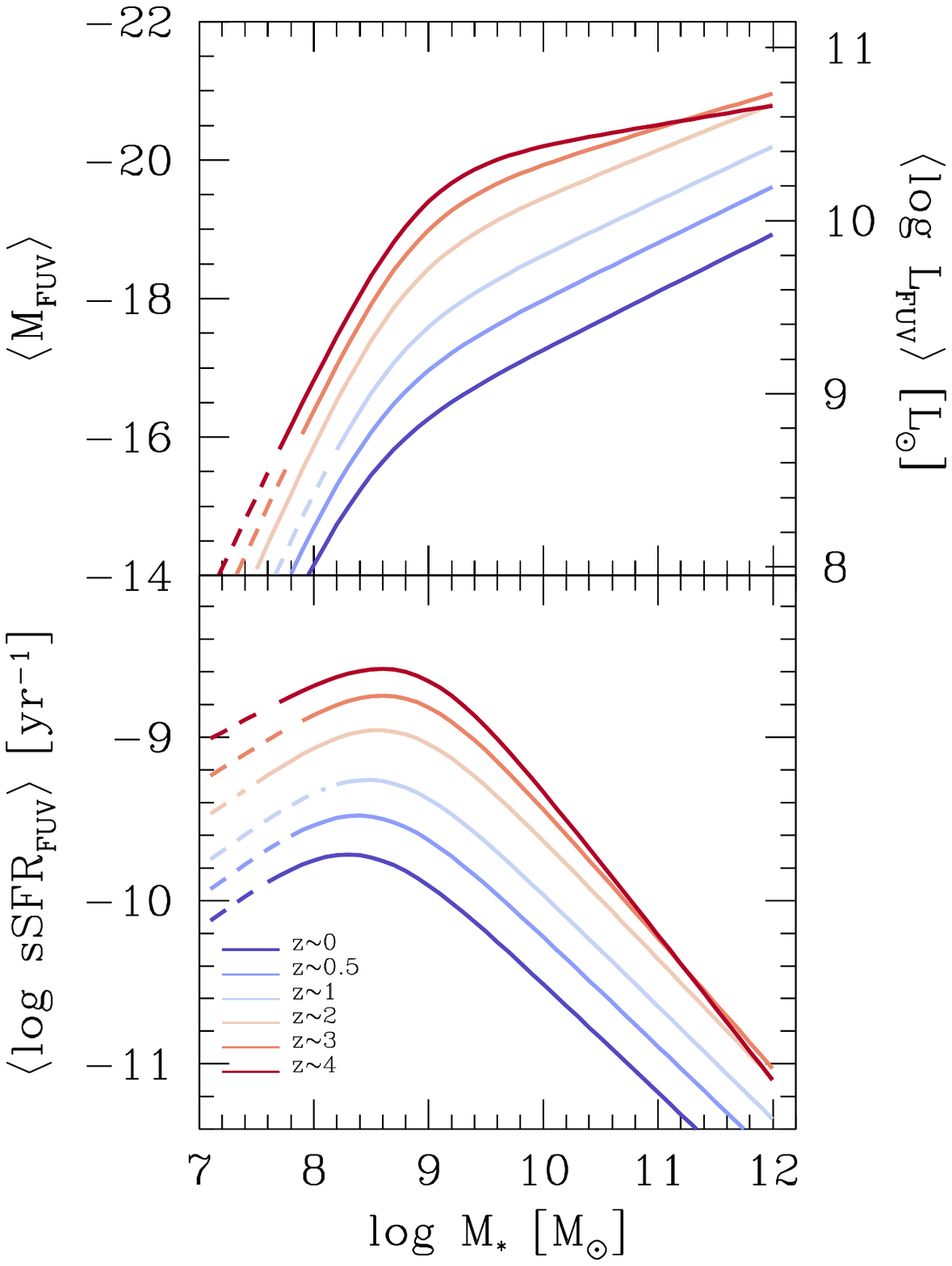}
	\vspace*{-120pt}
		\caption{{\bf Upper Panel:} The M$_{\rm FUV}$--\ms\ relation for MS SFG  from $z\sim0$ to $z\sim4$.
		The solid lines show our results constrained by our best fitting models of  
		the FUV LF. The dashed lines show the mass regimes
		when the FUV LFs have been extrapolated. 
		{\bf Bottom Panel}:The sSFR$_{\rm FUV}$--\ms\ relation for MS SFGs from $z\sim0$ to $z\sim4$.
		Notice the bend in the sSFR$_{\rm FUV}$--\ms
		relationship corresponds to the knee of the FUV LF, Figure \ref{fig:uv_lf}.
 	}
	\label{fig:ssfr_uv_vs_ms}
\end{figure}

Figure \ref{fig:ssfr_vs_z} compares our results to several observational studies listed 
 in Table \ref{tab:T3} for galaxies with masses above $\ms\sim0.5-1\times10^{9}\msun$. 
 The compiled data have been homogenized to a same \citet{Chabrier2003} IMF. In addition,
 in cases where the authors report the mean $\langle \sfr\rangle$ we use Equation (\ref{ec:mean_diff}) to transform into the mean
 $\langle\log \sfr\rangle$. We also compute the mean \sfr--\ms\ relationship at $z\sim0$ for SDSS galaxies 
 using the GALEX-SDSS-WISE Legacy
 catalogue \citep[GSWLC,][]{Salim+2016,Salim+2018}. Here, we use their deep GSWLC-2 catalogue based on UV, optical, and
 22$\mu$m data from WISE to derive SFRs using SED fitting techniques, for details see \citet{Salim+2018}. 
 From this catalog, we use their derived SFRs, based on both the UV and IR components, to 
 compute the average \sfr--\ms\ for  MS SFGs as described next. Firstly,  we note that we use a slightly modified version of
 the  \sfr--\ms\ relation by \citet{Speagle+2014} adapted for the GSWLC-2 catalogue. Then, galaxies
 $-0.7$ dex ($\sim2.3\sigma$) away from the ridgeline are excluded. Next, we performed a power-law fit to the MS for 
 the trimmed sample and again galaxies  $-0.7$ dex away from the ridgeline were excluded. We repeated the above processes
 three more times. We note that on the last iteration the parameters of the power-law fits did not change considerably.
 This is similar to what has been proposed in \citet{Fang+2018} in order to obtain relations that are close to the highest-density
 ridgeline of the MS. In Figure  \ref{fig:ssfr_vs_z}  the colors indicate the different calibrators
reported by the authors. Note, however, that most of the data compiled in this 
paper are based on FUV+IR measurements,
for a more direct comparison to our results. 

In excellent agreement with the set
of independent observations from Table \ref{tab:T3},
our results reproduce both the strong evolution of the \sfr\  with $z$ 
and the dependence on \ms. Even when we have extrapolated 
our best fits to the IR LFs, as indicated
by the dashed lines in Figure \ref{fig:ssfr_vs_z}, our results are consistent 
with the set of observations from Table \ref{tab:T3}. We emphasize that we did not fit 
to these data but our results fell out {\it naturally} from the the join fit to the LFs and the GSMF.

Figure \ref{fig:ssfr_vs_ms} shows the obtained 
\ssfr-\ms\ relation for MS galaxies from $z\sim0$ to $z\sim4$. Here, our results are again 
consistent with the compiled data. The consistency presented in Figures \ref{fig:ssfr_vs_z} and
\ref{fig:ssfr_vs_ms} is not trivial since the compilation described above is composed
of measurements based on different surveys and calibrators (though mostly are based on FUV+IR measurements).
We thus conclude that our derived $\sfr$s describe quite well a set of independent observational inferences over the range where
both the IR and FUV LFs are available. A straightforward implication from the above
is that the joint evolution of the GMSF and the FUV and IR LFs is self-consistent, that is, 
the evolution of the FUV+IR SFR of MS galaxies is consistent with their stellar mass growth. The reason for this 
is that in our approach the FUV+IR LF is the result of convolving the GSMF of SFGs with the 
SFR conditional distribution given \ms, see Equation (\ref{ec:phi_conv}) and Figure \ref{fig:cartoon}.

We notice that our derived sSFR-\ms\ relations at all redshifts are not well described by simple power laws.
 A curvature in the MS has been already documented in previous works \citep[see e.g.,][]{Whitaker+2014,Gavazzi+2015,Lee+2015,Schreiber+2015,Tomczak+2016,Lee+2018,Popesso+2019a,Leslie+2020}.
 In general, our MS at any $z$ follows a power law only at the high-mass end, with the mass at which 
 the relation departs from the power law decreasing with $z$, roughly, from $\ms\sim 10^{11}$ at $z>2$ 
 to $\ms\sim 2\times 10^{10}$ at $\sim 0$. The slope at $z\sim0$ for galaxies above $\ms\sim10^{10}\msun$ is 
 $\sim-0.71$. At lower masses and at any redshift, the MS bends until an inflection point
is attained. In the sSFR--\ms\ plane, this means that the MS changes sign. At $z\sim0$ we find that at low masses
$\ms\lesssim10^{8.8}\msun$ the slope is $\sim0.37$ while for $\ms\sim10^9-10^{10}\msun$ the slope is $\sim-0.37$. 
This is an interesting feature of the MS at low-masses that our method has allowed us to establish. 
Following, we discuss it in more detail. 

\subsection{The Main Sequence of low-mass and dwarf galaxies}

\begin{figure}
	\vspace*{-200pt}
	\hspace*{-20pt}
	\includegraphics[height=7in,width=5.8in]{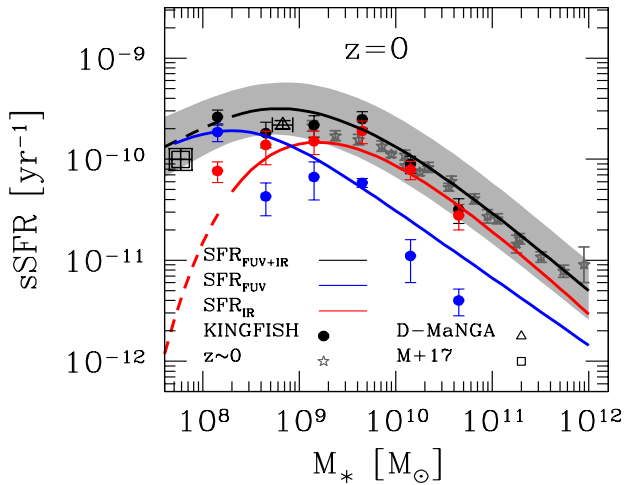}
	\vspace*{-100pt}
		\caption{The \ssfr-\ms\ relation for MS SFGs at $z\sim0$ 
		over a broad mass range: $\ms \sim7\times10^{7} - 10^{12}\msun$, black solid line.
		The contributions from FUV and IR are shown separately with blue and red solid 
		lines respectively. The gray stars with error bars correspond to the compilation 
		from Table \ref{tab:T3}. The solid black circles corresponds
		to the total SFRs based on the KINGFISH sample from \citet{Skibba+2011}, 
		while the blue and red solid circles correspond to the contributions from FUV and IR respectively. 
		The open triangle shows the average \ssfr\ from a sample of low-mass galaxies at the mass range 	
		$10^{8.5}\lesssim \ms/\msun\lesssim 10^{9}$
		from the MaNGA/SDSS-IV survey (Cano-Diaz in prep.). The open square shows the results 
		from  a sample of dwarf galaxies from \citet{McGaugh+2017} at $10^{7.5}\lesssim \ms/\msun\lesssim 10^{8}$.
		Our results are consistent with this set of independent observations for local galaxies.  
 	}
	\label{fig:ssfr_vs_ms_local}
\end{figure}

Next, we explore our results for low-mass galaxies
$10^{8}\lesssim\ms/\msun\lesssim10^{9}$. 
Our low-mass limits are restricted by the observational availability of both the FUV and IR LFs. This 
affects the robustness of our derivations at masses below $\ms\sim5\times10^9$ \msun, 
in particular for redshifts larger than $z\sim0.5$. 
For low-mass galaxies, our results are based mainly on the FUV 
LFs, and to a lesser extent, on the extrapolations of the IR LFs, 
due to the flat faint-end slopes of the latter (see Figure \ref{fig:ir_lf}). 
It is thus important to understand the robustness of our inferences for low-mass galaxies and the impact of 
our extrapolations from the IR LFs. 

We begin by noticing that 
previous authors reported a strong dependence of dust-obscured star formation with stellar mass as follows:
the SFRs of galaxies with $\ms\gtrsim10^9\msun$ 
are mainly traced by the IR light while the FUV light is more important for
lower mass galaxies, and this trend is consistent with no redshift dependence \citep[e.g.,][]{Pannella+2009,Whitaker+2017}. 
If we extrapolate the above results to our derivations, then the implication is straightforward:
for low-mass galaxies $\ms\lesssim10^9\msun$, the FUV LF 
is more important for the total SFR than the IR LF. In other words, little or no dust corrections are required for the FUV-based
SFRs of low-mass galaxies. In consequence, our resulting trends with mass and redshift for  
low-mass galaxies, that are dominated by FUV, will be in the right direction. The above is also supported by
Figures \ref{fig:sfr_vs_z} and \ref{fig:fraction_obsc} below.  
In the light of the above, next we describe our results for low-mass galaxies. 

Figure \ref{fig:ssfr_vs_ms} shows evidence that the \ssfr--\ms\ relation 
for MS galaxies {\it bends}, and even the slope sign changes, 
at the low-mass end at all redshifts. Why does the \ssfr--\ms\ relation change sign at the low-mass end? 
To answer this question, we use the fact that the SFRs from low-mass galaxies, 
$\ms\lesssim10^{9}\msun$, are dominated by the FUV component. 
The upper and bottom panels of Figure \ref{fig:ssfr_uv_vs_ms} show the corresponding
M$_{\rm FUV}$--\ms\ and sSFR$_{\rm FUV}$--\ms\ relationships. An important trend is apparent. The 
sSFR$_{\rm FUV}$--\ms\  relation bends below $\ms\sim10^{9}\msun$, similarly to the  \ssfr--\ms\ relation.
Not surprising, this bend is inherited by the turnover of the M$_{\rm FUV}$--\ms\ relation. The next 
important feature to notice is that the turnover mass
of the M$_{\rm FUV}$--\ms\ relation is nearly constant with redshift, $\sim5\times10^{8}\msun$, but
M$_{\rm FUV}$ changes from M$_{\rm FUV}\sim-16$ at $z\sim0$ to M$_{\rm FUV}\sim-20$ at $z\sim4$.
By looking to Figure \ref{fig:uv_lf}, it is then evident that the above magnitudes correspond to the {\it knee} of the FUV LF.
It is now clear that the bending in the \ssfr--\ms\ is due the knee of the FUV LF. In general, it is important to note
that the form of the M$_{\rm FUV}$--\ms\ and sSFR$_{\rm FUV}$--\ms\ relations
are governed by the Schechter-like shapes of the FUV LF.
Individual determinations for low-mass galaxies at high redshifts will be key to confirm the
above and for studying the self-consistency of the joint evolution of the faint end of the GSMF and the FUV+IR LFs. 
Fortunately, more accurate observational constraints are available  at low-masses for nearby galaxies,
so, we now focus our discussion on low-mass and dwarf galaxies at $z\sim0$. 

Figure \ref{fig:ssfr_vs_ms_local} presents again the  \ssfr--\ms\ relationship but this time  
at $z\sim0$ and extending down to dwarf
galaxies, $\ms\sim7\times10^{7}\msun$. As noted above, the slope of this 
relation flattens at low-masses and changes sign 
around $\ms\sim5\times10^{8}\msun$. Additionally, we present separately the contribution to the SFRs
from FUV, blue solid line, and IR, red solid line.
Notice that the mass around which the slope of the \ssfr--\ms\ relationship changes sign 
is close to the mass above which the regime of dust-obscured 
star formation starts to dominate, $\ms\sim6\times 10^{8}\msun$. 

Similarly to Figure \ref{fig:ssfr_vs_ms}, we compare our results with our compilation 
at $z\sim0$, gray stars with error bars. At low-masses, we add our estimates of the
 (total) sSFRs to the KINGFISH sample from \citet{Skibba+2011} 
obtained with GALEX and the Herschel Space Observatory data, filled black circles. 
We also compute the average \ssfr\ from a sample of low-mass galaxies at the mass range $10^{8.5}\lesssim \ms/\msun\lesssim 10^{9}$
with an axis ratio of $b/a\geq0.5$
from the MaNGA/SDSS-IV survey (Cano-Diaz in prep.).\footnote{The methodology to derive the SFRs and to select the SF galaxies are both explained in detail in \citet{CanoDiaz+2019}. In this work we use the data products provided by the Pipe3D Value Added Catalogue \citep{Sanchez+2018}, which uses the Integral Field Spectroscopy analysis pipeline,Pipe3D \citep{Sanchez+2016}, for the last MaNGA public data release: version v2-4-3.} Finally, from a sample of dwarf galaxies
from \citet[][]{McGaugh+2017} we compute the average \ssfr\ at the mass range $10^{7.5}\lesssim \ms/\msun\lesssim 10^{8}$, open square.
Notice that, both for the MaNGA and the \citet{McGaugh+2017} samples, the SFRs were derived from H$_\alpha$ luminosities. We transform their H$_\alpha$ determinations
into dust-corrected \sfruv\ by using eq. (16) from \citet{Shin+2019}.
In general, our results for the total $\sfr$s are in good agreement
with the above set of observations down to $\ms\sim7\times10^{7}\msun$. 

\begin{figure}
	\vspace*{-80pt}
	\includegraphics[height=9.5in,width=7.8in]{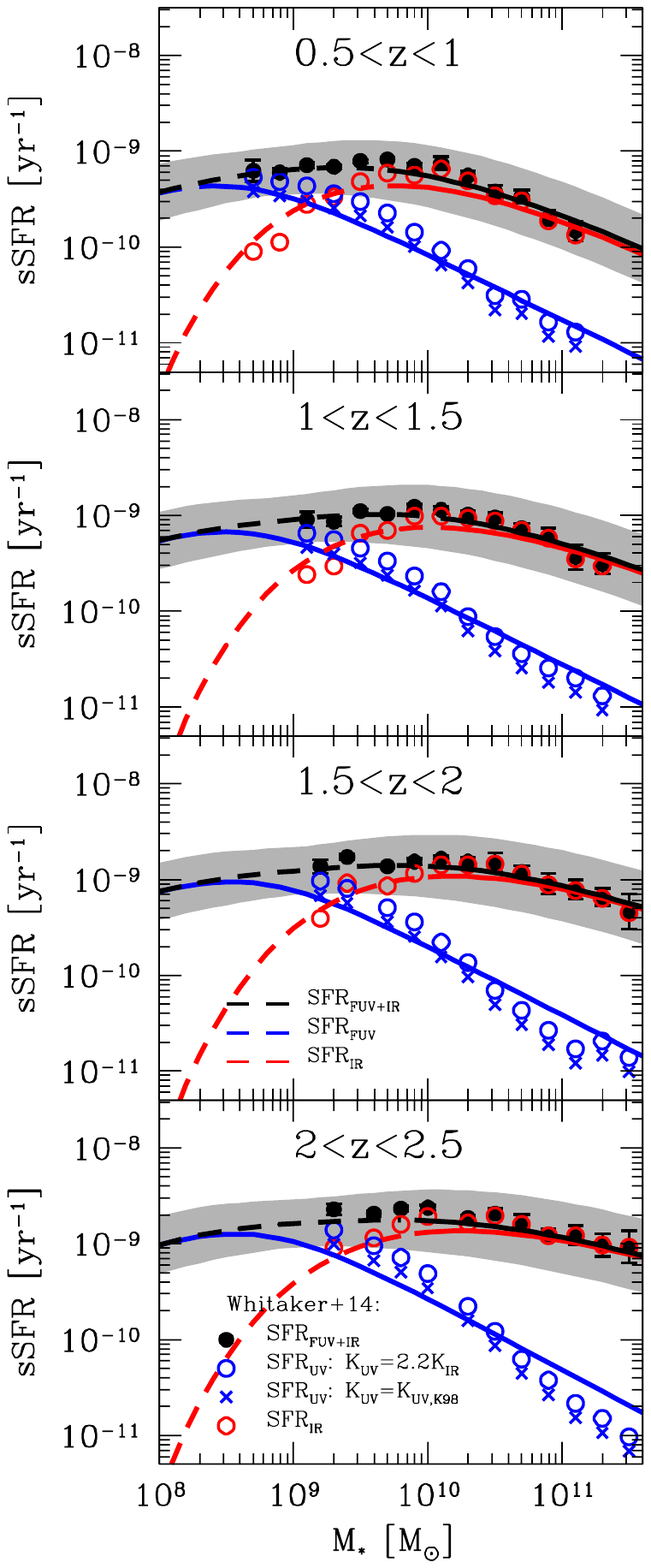}
	\vspace*{-140pt}
	\caption{The predicted total average \ssfr-\ms\ relation and the contribution from FUV and IR 
		light from $z=0.5$ to $z=2.5$.
		We compare to the inferences from \citet{Whitaker+2014}. Similarly to Figure \ref{fig:ssfr_vs_ms}, the solid lines
		show our resulting sSFRs where both the best fitting models
		from the FUV and IR LFs are valid while the dashed lines show where the FUV LF is valid
		but the IR LF has been extrapolated. Our results are consistent with the total SFRs
		from \citet{Whitaker+2014} as well as for the FUV and IR contributions, even in the regimes
		where we have extrapolated our results.}
	\label{fig:sfr_vs_z}
\end{figure}

In the case of the KINGFISH sample, the authors reported IR luminosities (see their table 1) which we transform into SFRs (red circles). We also
subtract IR SFRs from the total in order to compute UV SFRs (blue circles) according to their equation 7. 
It is encouraging that our results reproduce pretty well the observational trends from the KINGFISH sample. 
Remarkably, both our results and the KINGFISH sample appear to have a similar characteristic mass, $\mobsc\sim6\times10^{8}\msun$,
above which the SFRs are dust-obscured. We thus confirm that the FUV light becomes more important as
a diagnostic of the SFR of dwarf galaxies. 

\begin{figure*}
	\vspace*{-250pt}
	\hspace*{30pt}
	\includegraphics[height=8.5in,width=6.2in]{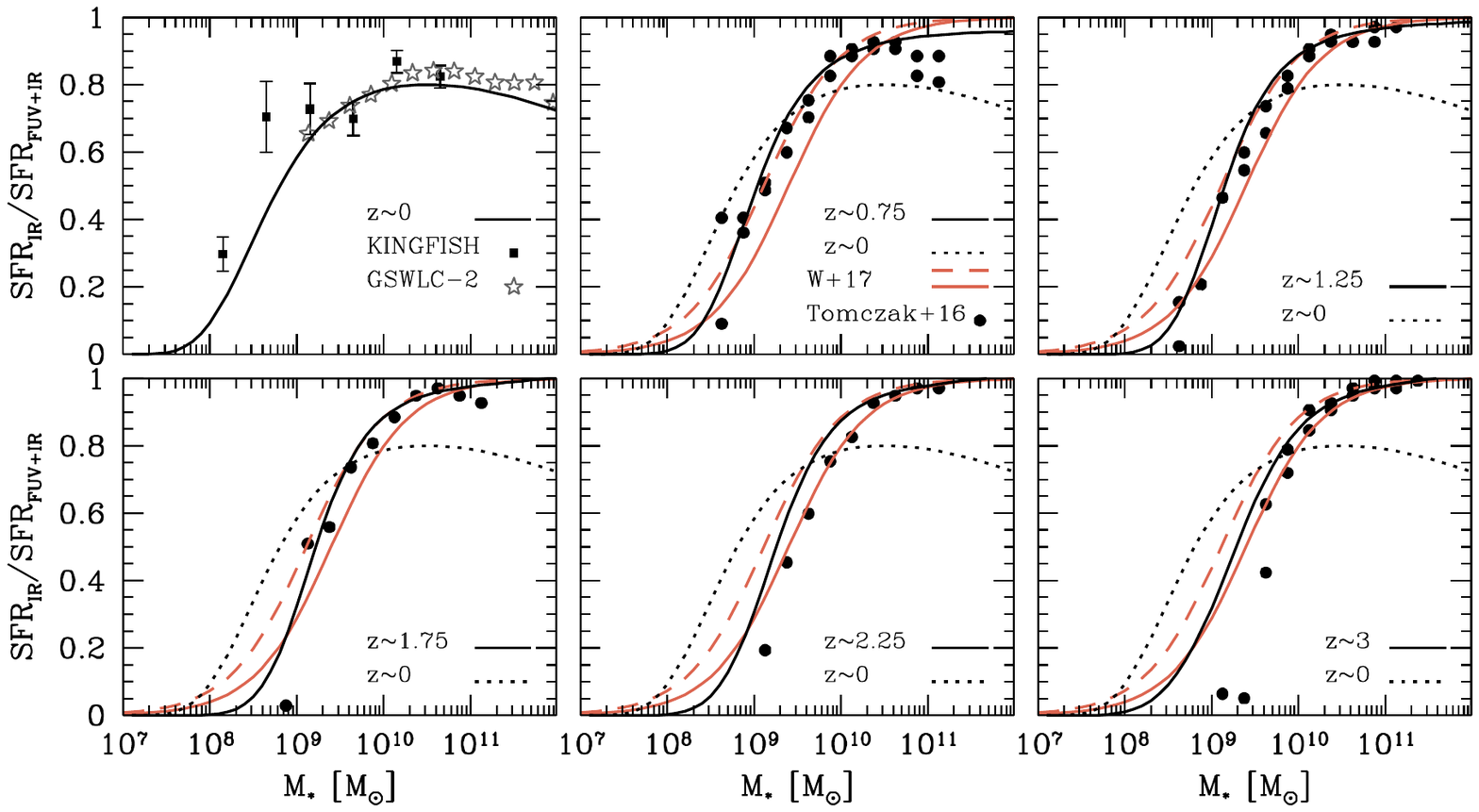}
	\vspace*{-130pt}
			\caption{Redshift evolution for the fraction of obscured star formation as a function of \ms. The
		solid lines show our results at the redshifts indicated by the labels while the dotted lines reproduce the 
		our results for local galaxies. It is clear that the fraction of dust obscured star formation evolves with 
		redshift. We present comparisons with the local determinations based on the KINGFISH sample 
		\citep[][filled black squares]{Skibba+2011}  and based on the 
		dust attenuation values $A_{\rm FUV}$ reported in the GSWLC-2 \citep[][gray stars]{Salim+2018}. At high
		redshift we reproduce the results from  \citet[][filled circles]{Tomczak+2016} based on their 
		reported luminosities $\langle L_{\rm FUV}\rangle$ and $\langle L_{\rm IR}\rangle$. 
		The red solid line is the best fit to observations from \citet{Whitaker+2017} 
		using the FUV estimator from \citet{Whitaker+2014} while the red dashed line show their results based on
		the \citet{Murphy+2011} FUV estimator.}
	\label{fig:fraction_obsc}
\end{figure*}

\subsection{The redshift evolution of the FUV and IR contribution to the total \sfr}

\begin{figure}
	\vspace*{-100pt}
	\hspace*{-15pt}
	\includegraphics[height=8in,width=5.8in]{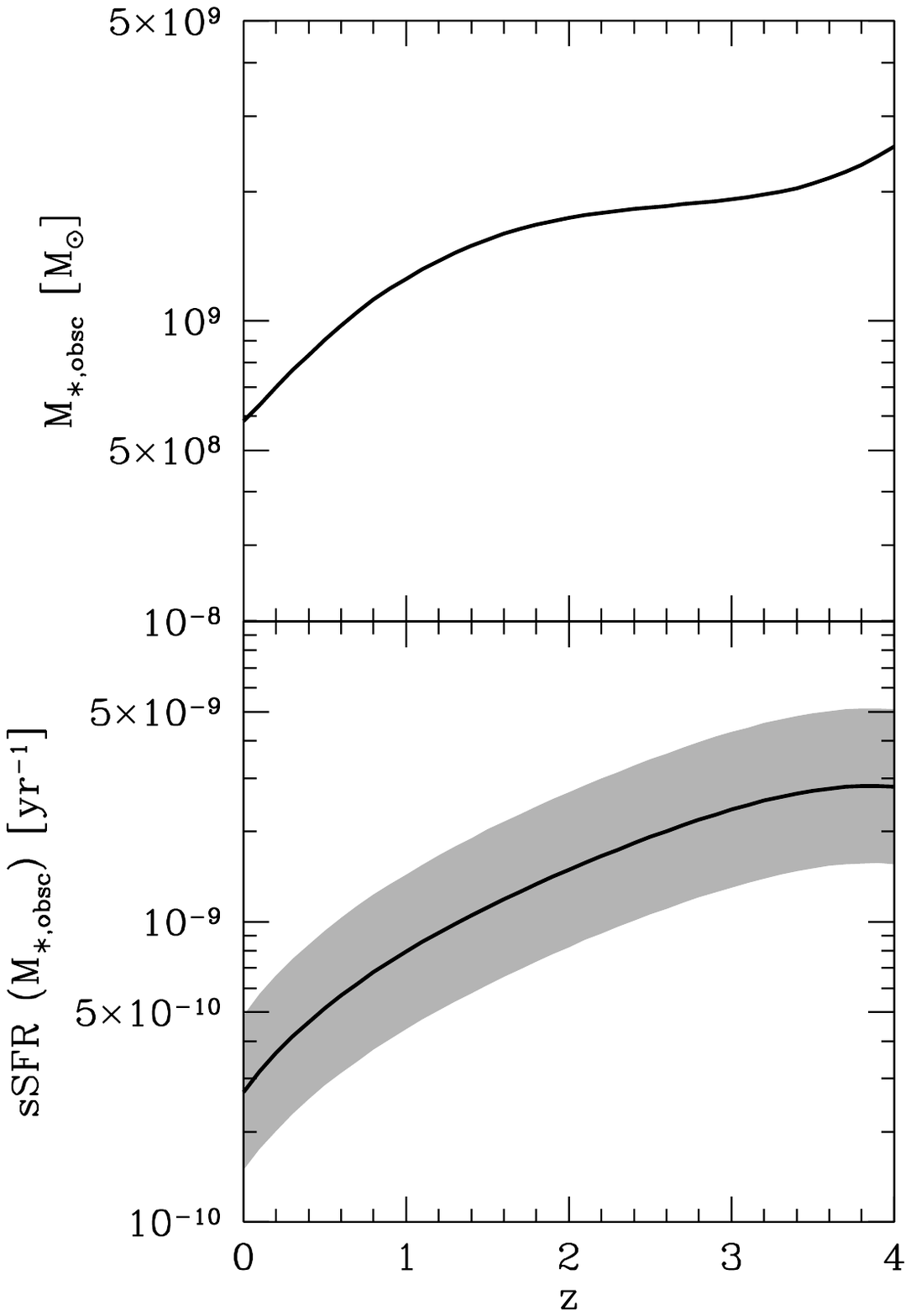}
	\vspace*{-135pt}
		\caption{ {\bf Upper panel:}  Redshift evolution of the characteristic mass \mobsc\  that marks the transition between the unobscured and obscured star formation regimes.
		Notice that since $z\sim1.5$ the characteristic mass \mobsc\  decreased by a factor of $\sim4$ while above $z\sim1.5$ has increased only slowly.   
		 {\bf Bottom panel:} Corresponding total \ssfr\ of the characteristic mass \mobsc. The $\ssfr$s has decreased an order of magnitude since $z\sim4$. 
		}
	\label{fig:mchar}
\end{figure}

In Figure \ref{fig:sfr_vs_z}, we explore the contribution of  FUV and IR  
to the total \ssfr--\ms\ relationship at $0.5\leq z\leq 2.5$. 
Similarly to Figure \ref{fig:ssfr_vs_ms}, the solid lines show the mass regime where the best fitting 
models of both the FUV and IR LFs are constrained
by the data while the dashed lines show the stellar mass regime where the LFs were extrapolated.
 
We compare our results to
the total SFRs reported from \citet{Whitaker+2014} computed as the combination of the UV\footnote{\citet{Whitaker+2014} 
estimated UV luminosities as the integrated light within the range between 1260--3000 \AA\ by using the 2800 \AA\ rest-frame luminosity: 
$L_{\rm UV}(1260-3000 \AA) = 1.5 \nu L_{\nu,2800}$.} 
and IR values based on the {\it Spitzer/} MIPS 24 $\mu$m photometry. 
Note that we subtract $\sim0.13$ dex to their 
SFRs in order to convert from $\log\langle{\rm sSFR}\rangle$ to $\langle\log \ssfr\rangle$, 
see Equation (\ref{ec:mean_diff}) and Table \ref{tab:T3}. 
In the same figure we present separately the 
contribution from FUV and IR by using the average luminosities, $\langle{L}\rangle(\ms)$, reported in their table 2. 
Note that we plotted two different symbols 
for the \sfruv\ from \citet{Whitaker+2014}. The open circles show the conversion factor utilized by 
\citet{Whitaker+2014} of $ \kfuv = 2.2\times \kir = 2.4 \times 10^{-10}\msun$ yr$^{-1}L_{\odot}^{-1}$ while 
the cross symbols use the conversion factor utilized in this paper, see Section \ref{sec:abundance_matching}.
The conversion factors used by \citet{Whitaker+2014} results in a factor of $\sim1.4$ larger than ours. 
Thus, the cross symbols are more adequate for comparing with our results.

In general, when comparing to \citet{Whitaker+2014} we observe that our results 
capture the same trends to the contributions from FUV and IR at all 
masses and redshifts, even for the regimes where we extrapolated our results from the
IR LFs, red dashed lines. Note that the FUV component dominates the low-mass regime both in our and in the
\citet{Whitaker+2014} results.
This is encouraging and hints again that the trends reported in 
Figures \ref{fig:ssfr_vs_ms} and \ref{fig:ssfr_vs_ms_local}
for the total SFRs at the low-mass regime are in the right direction.

Regarding the characteristic mass, \mobsc, where the \sfruv\ and \sfrlir\ are equal, or equivalently, 
the transition mass to the dust-obscured star formation regime, we observe an increase
in mass by a factor of  $\sim3.3$
from $z\sim0$ to $z\sim2.5$; the values are respectively $\mobsc\sim6\times 10^{8}\msun$ and 
$\mobsc\sim 2\times 10^{9}\msun$. From $z\sim0.5$ to $z\sim2.5$ 
the characteristic mass has increased by a factor of $\sim 2.2$. A similar trend is observed based
on the results from \citet{Whitaker+2014} and reported explicitly in \citet{Whitaker+2017}.  

\subsection{Mass and redshift dependence of the obscured star formation regime}

According to Figures \ref{fig:ssfr_vs_ms_local} and \ref{fig:sfr_vs_z}, 
our results are consistent with observational samples that 
separate the contributions from FUV and IR from $z\sim0$ to $z\sim2.5$, even when we 
extrapolated our results for the IR LFs. 
In the following we will use our results based on the extrapolations of the IR LFs as direct predictions 
from our approach. 

The upper left panel of Figure \ref{fig:fraction_obsc}  presents our resulting fraction of dust-obscured 
star formation as a function of stellar mass at $z\sim0$, black solid line. 
Our results 
show a strong dependence with stellar mass 
in the direction that the SFRs of high-mass galaxies are more obscured by dust than low-mass
galaxies. At $\ms\sim10^{10}\msun$ the dust-obscured fraction approaches 
to a maximum value of  $\sim0.8$ and for larger masses it keeps roughly constant. 
The filled circles show the results from the KINGFISH sample. The gray stars show the results of using the 
dust attenuation values $A_{\rm FUV}$ reported in the GSWLC-2 catalog. 
Both, the KINGFISH sample and  GSWLC-2 catalog are in good agreement with our results. 
Next, we discuss the comparison with high redshift data. 

The remaining panels of Figure \ref{fig:fraction_obsc} present
the fraction of dust-obscured star formation as a function of stellar mass from $z\sim0.75$ to $z\sim3$. 
In each panel the $z\sim0$ curve is repeated with the dotted line. As seen from the
sequence of panels, 
the fraction of the dust-obscured SFRs 
evolves mostly from $z\sim0$ to $z\sim1.2$ but above $z\sim1.2$ our results are consistent 
with little evolution up to $z\sim3$. Thus our results indicate that the fraction of dust-obscured SFRs evolves
with redshift. We also reproduce the best fitting model to the fraction of dust-obscured SFRs 
from \citet{Whitaker+2017} at $0.5\leq z\leq 2.5$ using their ``standard calibration", black solid line, and their best fitting 
model based on the \citet{Murphy+2011} calibration parameters, dashed lines. 
Notice the agreement between our results with those from \citet{Whitaker+2017} when considering
both of their calibrators. We also compute
the fraction of dust-obscured SFRs using the luminosities $\langle L_{\rm FUV}\rangle$ and $\langle L_{\rm IR}\rangle$ 
reported in table 1 from \citet{Tomczak+2016} as filled circles. We find that our results are consistent with these authors, 
including the increase in \mobsc\ with redshift.

An important feature is worth of mentioning here. 
According to our results, at low redshifts, $z\lesssim0.75$, high-mass galaxies have become 
more ``transparent"  to UV light compared to their high redshift counterparts. For low- and 
intermediate-mass galaxies we observe the opposite, they have 
become more obscured by dust. Similar 
trends are derived when using the  \citet{Tomczak+2016} dust-obscured fractions. 

The top panel of Figure \ref{fig:mchar} presents the evolution of the characteristic mass, \mobsc, at 
which the fraction  of dust-obscured star formation is 0.5 based on our model.  
Here is evident the rapid evolution of \mobsc\
for, $z\lesssim1.2$, but then it rises only slowly at high redshifts up to a mass of $\mobsc\sim2\times10^{9}\msun$, as noted previously.  
The bottom panel shows the corresponding \ssfr\ of \mobsc.
The \ssfr(\mobsc) has a strong correlation with $z$ and appears to reach a maximum 
of $\ssfr\sim 3\times 10^{-9}$ yr$^{-1}$ just before $z\sim4$.
The \ssfr(\mobsc)-redshift relation has decreased by an order of magnitude since $z\sim4$.

\section{Summary  and Discussion} 
\label{sec:discussion}

In this paper we present an analytical method for deriving the evolution of the mean \ssfr--\ms\ relation 
of MS galaxies by combining the FUV and IR rest-frame LFs with the 
GSMF of SFGs from $z\sim 0$ to $z\sim4$. The total SFR is estimated as the sum of the unobscured and obscured
regimes traced by the FUV ($1500\AA$) and IR ($8-1000\mu$m) luminosities, respectively.
Our approach is an alternative to the commonly employed procedure in the literature, 
consisting on using large galaxy samples for which the masses and SFRs are inferred invidually for every galaxy.
As discussed in Section \ref{sec:abundance_matching}, the determination and understanding  
of the \sfr--\ms\ relation and its evolution for MS galaxies can be affected 
due to {\it i)} intrinsic selection effects in large galaxy samples,
and {\it ii)} the comparison of SFRs based on different 
tracers and estimations is not trivial. In our approach, this is not the case since the evolution of the
MS is derived from the FUV and IR LFs and
the GSMF, distributions that are complete in volume, luminosities and mass over a
large redshift range. Our main results are as follows:

\begin{itemize}

\item The homogeneously obtained SFRs as a function of \ms\ 
for main sequence, MS, galaxies at redshifts $0\lesssim z\lesssim4$ confirm and unify,
within the scatter, the values obtained previously from different galaxy surveys 
that used a diversity of SFR tracers and methods, see Figure  \ref{fig:ssfr_vs_z}. 

\item Our methodology allows for consistent inferences of the mean \ssfr--\ms\ relationships for MS galaxies 
down to low masses. Moreover, it allows for reasonable extrapolations down to the regime of dwarf 
galaxies, where the unobscured FUV component dominates the total SFR, see Figures \ref{fig:ssfr_vs_ms} and 
\ref{fig:ssfr_vs_ms_local}.


\item The mean \ssfr--\ms\ relation for MS galaxies bends strongly at low-masses, $\lesssim10^{9}\msun$, 
and the slope sign changes
from positive at low-masses to negative at intermediate and high masses  masses at all redshifts. In 
particular at $z\sim0$ the slope change between $\sim0.37$ to $\sim-0.37$ at low to intermediate masses. 
The bending in the MS at lower masses is connected to the knee of the FUV LF, see Figure \ref{fig:ssfr_uv_vs_ms}. 

\item At $z\sim0$ the change of sign in the MS
occurs around $\ms\sim5\times10^{8}\msun$ which is close
to the dust-obscured star formation regime at this redshift, $\ms\sim6\times10^{8}\msun$. 

\item  At $z\sim0$, the resulting contributions from unobscured, FUV, and obscured, IR, SFRs are 
in good agreement with the results from the KINGFISH sample (based on GALEX and Herschel
Space Observatory data) and the dust attenuation values $A_{\rm FUV}$ reported in the 
GSWLC-2 (based on GALEX, SDSS and WISE). At $0.5\lesssim z\lesssim2.5$ we 
show that our results capture 
the observed trends to the contributions from FUV and IR to the total SFRs from \citet{Whitaker+2014}. 

\item At all redshifts the contribution from FUV dominates the total  \ssfr--\ms\ relation  
at low-masses, Figures \ref{fig:sfr_vs_z} and \ref{fig:fraction_obsc}. 
The characteristic mass for the obscured SFR regime has decreased a factor
of $\sim3$ since $z\sim4$: $M_{\rm *,obsc}\sim 2\times10^{9}\msun$ at $z\sim4$ and 
$M_{\rm *,obsc}\sim 6\times10^{8}\msun$ at $z\sim0$, see Figure \ref{fig:mchar}.

\item The fraction of obscured SFR, SFR$_{\rm IR}/$SFR$_{\rm FUV+IR}$, depends strongly on 
mass and it changes very little with redshift  for $z>1.2$, in agreement with 
\citealp{Tomczak+2016,Whitaker+2017},  
Figure \ref{fig:fraction_obsc}. Below $z\sim0.75$,  galaxies more massive than $\sim 10^{10}$ \msun\ 
become more ``transparent" than their high-redshift galaxies with the same stellar mass, while for 
low-mass galaxies, the opposite is true as they become more obscured 
by dust  at the same stellar mass, see Figure \ref{fig:fraction_obsc}.

\end{itemize}
 
According to our results, the \ssfr--\ms\ relation bends and changes its slope sign 
from positive at low masses to negative at high masses. At
$z\sim 0$, this change occurs at $\ms\sim 5\times 10^8\msun$. This has 
interesting implications for dwarf galaxies. If we interpret the inverse of the total \ssfr\ as the
characteristic time that it will take a galaxy to double its mass at a constant \sfr, then the above
implies that dwarf MS galaxies, $\ms\lesssim 5\times 10^8\msun$, form stars at a lower pace 
than intermediate-mass MS galaxies, $10^9\lesssim\ms/\msun\lesssim10^{10}$. For our low-mass limit of $\ms\sim7\times10^{7}\msun$,
we find that $\ssfr\sim10^{-10}$ yr$^{-1}$, see Figure \ref{fig:ssfr_vs_ms_local}. 
Assuming a constant SFH, this implies
an assembling time of $\sim 10$ Gyrs. Observations of nearby dwarf galaxies show a significant mass 
fraction, $\gtrsim 30$\%, in stellar populations older than 10-11 Gyr \citep{Weisz+2014}, consistent with 
our estimation. Note that if the \ssfr\ of low-mass galaxies would follow 
the same trend with \ms\ as the one for $\ms\gtrsim 10^{9}\msun$, then the assembling time 
would be of $\lesssim 2$ Gyr. That is, dwarf galaxies would be just in their process of
formation, in clear disagreement with our results and with the resolved 
SFHs from nearby dwarf galaxies.  

There are other observational works of local star-forming dwarf 
galaxies that also show that their mean \ssfr-\ms\ relation bends with respect to the
MS at higher masses \citep[e.g.,][]{McGaugh+2017,Davies+2019}. However, 
in these works the slope of the low-mass MS is not as steep as in our case, though this is difficult 
to evaluate since the scatter of the MS increases at lower masses 
\citep[see e.g.,][]{Davies+2019}. Motivated by the above, we have repeated our calculations at $z\sim0$
by increasing the scatter only for the FUV contribution from 0.3 dex to 0.5 dex (recall  
that for low-mass galaxies the FUV is more important). We find that the slope at low-masses of the resulting 
mean \ssfr--\ms\ relation becomes slightly shallower as compared to the results based on the dispersion of 0.3 dex. 
More accurate observational studies of low-mass and in particular dwarf galaxies will be key to 
confirm the bend of the mean \ssfr--\ms\ relation.

Studying the joint evolution of the  FUV and IR rest-frame 
LFs and the GSMF appears as a promising and powerful approach to understand the contribution of unobscured (FUV) and
obscured (IR) to the total SFRs as a function \ms\ and $z$, and ultimately to constrain models of the formation and
destruction of dust in galaxies. 

The increase of heavy element abundances in any galaxy, and hence of its dust content, 
is a direct result of stellar activity. As stated by several authors
(see discussions by \citealp{Dwek+2011} and \citealp{Slavin+2020} and references therein), the 
expanding ejecta from supernovae is perhaps the main source of dust at high redshifts. Our results 
suggest that there seems to be a feedback mechanism between the formation of dust and the star 
forming activity. For high-mass galaxies at their early stages of evolution, the vigorous star forming 
activity supplies the ambient interstellar medium with a generous amount of recently formed dust. 
This may provide the required opacity to shield the neighbouring gas from the stellar UV radiation  
\citep[e.g.,][]{Franco+1986}, stimulating the formation of new generations of molecular clouds. This in turn 
induces the conditions for further star formation but the strong energy injection, both radiative and mechanical, 
is highly disruptive and tends to destroy the clouds, restricting the outcome of the star formation process. 
Regardless of the evolution of the resulting star formation rate, the emerging luminosity proceeds in a 
dust-obscured fashion. For the case of low mass galaxies, in contrast, with a smaller star formation rate 
and a weaker gravitational field than their massive counterparts, the evolution proceeds at a slower rate 
with a milder supernova rate and a slowly growing metal content. The stellar energy injection in this case 
can be more disruptive and even expell the interstellar gas along with the recently formed dust out of the 
galaxy \citep[e.g.,][]{Caproni+2015}, leading to a more transparent evolutionary mode. It is 
not clear, however, why the dust-obscured fraction evolves mostly below $z\sim1$ and 
there is little evolution between $z\sim1.5$ and $z\sim4$. Perhaps our results indicate that dust cycling 
is not the full story but the spatial distribution of dust 
and its dependence on disk secular evolution might play a relevant role  \citep[e.g.,][]{Dalcanton+2004}. 
These scenarios will be explored in more detail in a future publication.

Future research should be focused on extending the consistency between the
GSMF and the evolution of FUV and IR LFs to lower mass and higher redshift regimes. 
Here we found
that this is the case at $z\sim4$ and for $10^9\lesssim\ms/\msun\lesssim10^{11.5}$, by being in agreement with the observed trends of the \ssfr--\ms\ relationship 
and the dust-obscured fraction results obtained directly from galaxy surveys. 
Extending the semi-empirical modelling like those presented in \citet{Rodriguez-Puebla+2017,Moster+2018,Tacchella+2018}
and \citet{Behroozi+2019} to include FUV and IR LFs looks very promising and timely to understand how dark matter halos
build their stellar mass and the evolution of dust in their host galaxies (Rodr\'iguez-Puebla et al. in prep.). 
In addition, more theoretical work 
will be needed in order to understand the role of the dust and the trends with the fraction of obscured SFR
derived here.

\section*{Acknowledgments} 
We thank the anonymous referee for a constructive report that helped to improve this paper.
ARP and VAR acknowledge support from UNAM PAPIIT grant IA104118 and from the CONACyT  `Ciencia Basica' 
grant 285721.
This project makes use of the MaNGA-Pipe3D dataproducts. We thank the IA-UNAM MaNGA team for creating this catalogue, and the CONACYT-180125 project for supporting them

\appendix

\section{The GSMF}
\label{secc:gsmf}

\begin{figure*}
	\vspace*{-120pt}
	\hspace*{80pt}
	\includegraphics[height=5in,width=3.8in]{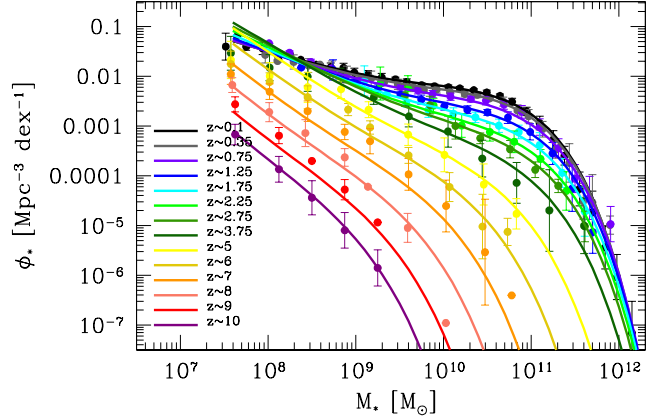}
	\vspace*{-60pt}
		\caption{Best fitting model to the redshift evolution of the GSMF from
		\citet{Rodriguez-Puebla+2017} .
 	}
	\label{fig:gsmf}
\end{figure*}

\begin{figure*}
	\vspace*{-250pt}
	\hspace*{30pt}
	\includegraphics[height=8.2in,width=6.2in]{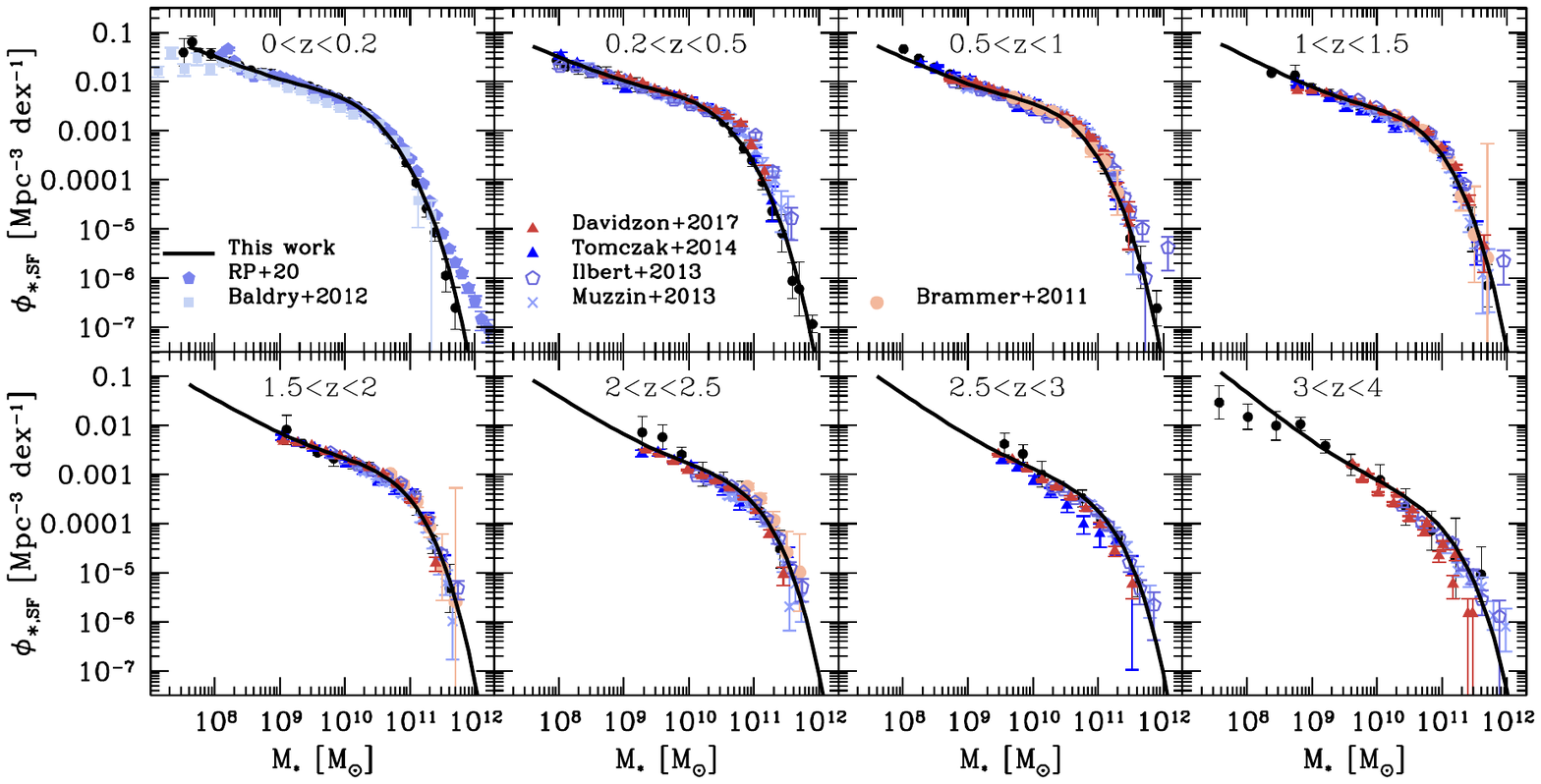}
	\vspace*{-130pt}
		\caption{ Galaxy stellar mass functions of SFGs at various redshifts. We compare
		our model for the GSMF of SFGs, solid line, with different authors based on local
		\citep{Baldry+2012,RP20} and high redshift \citep{Brammer+2011,Muzzin+2013,Ilbert+2013,Tomczak+2014,Davidzon+2017}  measurements. In general we observe a good agreement between the different authors and our model, however, the comparison should
		be taken with care as the definition of SFG could vary between authors. 
 	}
	\label{fig:gsmf_sf}
\end{figure*}

We define the GSMF for all (including SF and quenched) galaxies as the sum of two modified Schechter components, $\phi_{\ast} = \phi_{\ast,1} +\phi_{\ast,2} $:
\begin{equation}
	\phi_{\ast,i}(\ms) =\phi^{\ast}_i \left( \frac{\ms}{M_i} \right)^{1+\alpha_i}\exp\left[-\left(\frac{\ms}{M_i}\right)^{\beta_i} \right],
\end{equation}
we assume that $\beta_1 = 1$, $\phi_2^{\ast} = 6\times \phi_1^{\ast}$, $M =M_1 = M_2$ and that $\alpha_2 = 1 + \alpha_1$. 
The above guarantees that the first component is a Schechter function while the second component is a modifed Schechter
component dominating the massive-end, $\ms \gtrsim M_i$, of the GSMF. Similarly to the LFs we integrate our mass
function over the redshift range that the GSMF is being observed: $\phi_{\rm obs} (z_i, z_f) = \int \phi_{\ast}(z) dV(z) \diagup \left(V_i - V_f \right)$. The best fitting models to the compilation from \citet{Rodriguez-Puebla+2017} are:
\begin{equation}
	\begin{array}{l}
		\alpha(z) = (-1.758\pm 0.026)    {\;} +  
		\log(\mathcal{S}(z; 0,2.815\pm0.297,2.085\pm0.187,0) ),
	\end{array}
\end{equation}

\begin{equation}
	\begin{array}{l}
		\log\left(\frac{\phi_1^{\ast}(z)}{ {\rm Mpc}^{-3}{\rm dex}^{-1}}\right) = (-3.161\pm 0.070)    {\;} +  
		 \log(\mathcal{S}(z; 0,1.472\pm0.183,2.538\pm0.252,0) ), 
	\end{array}
\end{equation}

\begin{equation}
	\begin{array}{l}
		\log \left(\frac{M(z)}{{\rm M}_{\odot}}\right) = (-10.204\pm 0.118)    {\;} +  
		 \log(\mathcal{S}(z; 5.529\pm0.229,6.961\pm0.423,-0.338\pm0.139)), 
	\end{array}
\end{equation}
and
\begin{equation}
	\begin{array}{l}
		\beta = 0.611\pm 0.037.
	\end{array}
\end{equation}
Here the function $\mathcal{S}$ is a double power-law that depends on $z$ and has three free parameters ($p_1,p_2,p_3$):
\begin{equation}
	\mathcal{S}(z;p_1,p_2,p_3) = 2 \left[ \left(\frac{1+z}{p_1}\right)^{p_2}  +  \left(\frac{1+z}{p_1}\right)^{p_4} \right]^{-1}.
\end{equation}
Notice that $p_1$ is equivalent to a characteristic redshift at which the function $\mathcal{S}$ transits
from the power law $p_2$ to $p_3$. In addition at $p_1 = 1+z$ then $\mathcal{S}=1$. 
Figure \ref{fig:gsmf} compares our the best fitting model with observations. 
Finally, in order to compute the GSMF of SFGs we use the fraction of 
SFGs from \citet{Rodriguez-Puebla+2017} described in their Section 4.4.
Figure \ref{fig:gsmf_sf} compares our resulting GSMF of SFGs to several measurements 
from literature.

\bibliography{Bibliography}

\begin{thebibliography}{}
\expandafter\ifx\csname natexlab\endcsname\relax\def\natexlab#1{#1}\fi
\providecommand{\url}[1]{\href{#1}{#1}}
\providecommand{\dodoi}[1]{doi:~\href{http://doi.org/#1}{\nolinkurl{#1}}}
\providecommand{\doeprint}[1]{\href{http://ascl.net/#1}{\nolinkurl{http://ascl.net/#1}}}
\providecommand{\doarXiv}[1]{\href{https://arxiv.org/abs/#1}{\nolinkurl{https://arxiv.org/abs/#1}}}

\bibitem[{{Alavi} {et~al.}(2014){Alavi}, {Siana}, {Richard}, {Stark},
  {Scarlata}, {Teplitz}, {Freeman}, {Dominguez}, {Rafelski}, {Robertson}, \&
  {Kewley}}]{Alavi+2014}
{Alavi}, A., {Siana}, B., {Richard}, J., {et~al.} 2014, \apj, 780, 143,
  \dodoi{10.1088/0004-637X/780/2/143}

\bibitem[{{Alavi} {et~al.}(2016){Alavi}, {Siana}, {Richard}, {Rafelski},
  {Jauzac}, {Limousin}, {Freeman}, {Scarlata}, {Robertson}, {Stark}, {Teplitz},
  \& {Desai}}]{Alavi+2016}
---. 2016, \apj, 832, 56, \dodoi{10.3847/0004-637X/832/1/56}

\bibitem[{{Arnouts} {et~al.}(2005){Arnouts}, {Schiminovich}, {Ilbert},
  {Tresse}, {Milliard}, {Treyer}, {Bardelli}, {Budavari}, {Wyder}, {Zucca}, {Le
  F{\`e}vre}, {Martin}, {Vettolani}, {Adami}, {Arnaboldi}, {Barlow}, {Bianchi},
  {Bolzonella}, {Bottini}, {Byun}, {Cappi}, {Charlot}, {Contini}, {Donas},
  {Forster}, {Foucaud}, {Franzetti}, {Friedman}, {Garilli}, {Gavignaud},
  {Guzzo}, {Heckman}, {Hoopes}, {Iovino}, {Jelinsky}, {Le Brun}, {Lee},
  {Maccagni}, {Madore}, {Malina}, {Marano}, {Marinoni}, {McCracken}, {Mazure},
  {Meneux}, {Merighi}, {Morrissey}, {Neff}, {Paltani}, {Pell{\`o}}, {Picat},
  {Pollo}, {Pozzetti}, {Radovich}, {Rich}, {Scaramella}, {Scodeggio},
  {Seibert}, {Siegmund}, {Small}, {Szalay}, {Welsh}, {Xu}, {Zamorani}, \&
  {Zanichelli}}]{Arnouts+2005}
{Arnouts}, S., {Schiminovich}, D., {Ilbert}, O., {et~al.} 2005, \apjl, 619,
  L43, \dodoi{10.1086/426733}

\bibitem[{{Atek} {et~al.}(2018){Atek}, {Richard}, {Kneib}, \&
  {Schaerer}}]{Atek+2018}
{Atek}, H., {Richard}, J., {Kneib}, J.-P., \& {Schaerer}, D. 2018, \mnras, 479,
  5184, \dodoi{10.1093/mnras/sty1820}

\bibitem[{{Baldry} {et~al.}(2012){Baldry}, {Driver}, {Loveday}, {Taylor},
  {Kelvin}, {Liske}, {Norberg}, {Robotham}, {Brough}, {Hopkins}, {Bamford},
  {Peacock}, {Bland-Hawthorn}, {Conselice}, {Croom}, {Jones}, {Parkinson},
  {Popescu}, {Prescott}, {Sharp}, \& {Tuffs}}]{Baldry+2012}
{Baldry}, I.~K., {Driver}, S.~P., {Loveday}, J., {et~al.} 2012, \mnras, 421,
  621, \dodoi{10.1111/j.1365-2966.2012.20340.x}

\bibitem[{{Behroozi} {et~al.}(2019){Behroozi}, {Wechsler}, {Hearin}, \&
  {Conroy}}]{Behroozi+2019}
{Behroozi}, P., {Wechsler}, R.~H., {Hearin}, A.~P., \& {Conroy}, C. 2019,
  \mnras, 488, 3143, \dodoi{10.1093/mnras/stz1182}

\bibitem[{{Bernhard} {et~al.}(2014){Bernhard}, {B{\'e}thermin}, {Sargent},
  {Buat}, {Mullaney}, {Pannella}, {Heinis}, \& {Daddi}}]{Bernhard+2014}
{Bernhard}, E., {B{\'e}thermin}, M., {Sargent}, M., {et~al.} 2014, \mnras, 442,
  509, \dodoi{10.1093/mnras/stu896}

\bibitem[{{B{\'e}thermin} {et~al.}(2012){B{\'e}thermin}, {Dor{\'e}}, \&
  {Lagache}}]{Bethermin+2012}
{B{\'e}thermin}, M., {Dor{\'e}}, O., \& {Lagache}, G. 2012, \aap, 537, L5,
  \dodoi{10.1051/0004-6361/201118607}

\bibitem[{{Bhatawdekar} {et~al.}(2019){Bhatawdekar}, {Conselice},
  {Margalef-Bentabol}, \& {Duncan}}]{Bhatawdekar+2019}
{Bhatawdekar}, R., {Conselice}, C.~J., {Margalef-Bentabol}, B., \& {Duncan}, K.
  2019, \mnras, 486, 3805, \dodoi{10.1093/mnras/stz866}

\bibitem[{{Bouch{\'e}} {et~al.}(2010){Bouch{\'e}}, {Dekel}, {Genzel}, {Genel},
  {Cresci}, {F{\"o}rster Schreiber}, {Shapiro}, {Davies}, \&
  {Tacconi}}]{Bouche+2010}
{Bouch{\'e}}, N., {Dekel}, A., {Genzel}, R., {et~al.} 2010, \apj, 718, 1001,
  \dodoi{10.1088/0004-637X/718/2/1001}

\bibitem[{{Bouwens} {et~al.}(2007){Bouwens}, {Illingworth}, {Franx}, \&
  {Ford}}]{Bouwens+2007}
{Bouwens}, R.~J., {Illingworth}, G.~D., {Franx}, M., \& {Ford}, H. 2007, \apj,
  670, 928, \dodoi{10.1086/521811}

\bibitem[{{Bouwens} {et~al.}(2011){Bouwens}, {Illingworth}, {Oesch},
  {Labb{\'e}}, {Trenti}, {van Dokkum}, {Franx}, {Stiavelli}, {Carollo},
  {Magee}, \& {Gonzalez}}]{Bouwens+2011}
{Bouwens}, R.~J., {Illingworth}, G.~D., {Oesch}, P.~A., {et~al.} 2011, \apj,
  737, 90, \dodoi{10.1088/0004-637X/737/2/90}

\bibitem[{{Bouwens} {et~al.}(2012){Bouwens}, {Illingworth}, {Oesch}, {Franx},
  {Labb{\'e}}, {Trenti}, {van Dokkum}, {Carollo}, {Gonz{\'a}lez}, {Smit}, \&
  {Magee}}]{Bouwens+2012}
---. 2012, \apj, 754, 83, \dodoi{10.1088/0004-637X/754/2/83}

\bibitem[{{Bouwens} {et~al.}(2015){Bouwens}, {Illingworth}, {Oesch}, {Trenti},
  {Labb{\'e}}, {Bradley}, {Carollo}, {van Dokkum}, {Gonzalez}, {Holwerda},
  {Franx}, {Spitler}, {Smit}, \& {Magee}}]{Bouwens+2015}
---. 2015, \apj, 803, 34, \dodoi{10.1088/0004-637X/803/1/34}

\bibitem[{{Brammer} {et~al.}(2011){Brammer}, {Whitaker}, {van Dokkum},
  {Marchesini}, {Franx}, {Kriek}, {Labb{\'e}}, {Lee}, {Muzzin}, {Quadri},
  {Rudnick}, \& {Williams}}]{Brammer+2011}
{Brammer}, G.~B., {Whitaker}, K.~E., {van Dokkum}, P.~G., {et~al.} 2011, \apj,
  739, 24, \dodoi{10.1088/0004-637X/739/1/24}

\bibitem[{{Brinchmann} {et~al.}(2004){Brinchmann}, {Charlot}, {White},
  {Tremonti}, {Kauffmann}, {Heckman}, \& {Brinkmann}}]{Brinchmann+2004}
{Brinchmann}, J., {Charlot}, S., {White}, S.~D.~M., {et~al.} 2004, \mnras, 351,
  1151, \dodoi{10.1111/j.1365-2966.2004.07881.x}

\bibitem[{{Burgarella} {et~al.}(2013){Burgarella}, {Buat}, {Gruppioni},
  {Cucciati}, {Heinis}, {Berta}, {B{\'e}thermin}, {Bock}, {Cooray}, {Dunlop},
  {Farrah}, {Franceschini}, {Le Floc'h}, {Lutz}, {Magnelli}, {Nordon},
  {Oliver}, {Page}, {Popesso}, {Pozzi}, {Riguccini}, {Vaccari}, \&
  {Viero}}]{Burgarella+2013}
{Burgarella}, D., {Buat}, V., {Gruppioni}, C., {et~al.} 2013, \aap, 554, A70,
  \dodoi{10.1051/0004-6361/201321651}

\bibitem[{{Cano-D{\'\i}az} {et~al.}(2019){Cano-D{\'\i}az}, {{\'A}vila-Reese},
  {S{\'a}nchez}, {Hern{\'a}ndez-Toledo}, {Rodr{\'\i}guez-Puebla}, {Boquien}, \&
  {Ibarra-Medel}}]{CanoDiaz+2019}
{Cano-D{\'\i}az}, M., {{\'A}vila-Reese}, V., {S{\'a}nchez}, S.~F., {et~al.}
  2019, \mnras, 488, 3929, \dodoi{10.1093/mnras/stz1894}

\bibitem[{{Cano-D{\'\i}az} {et~al.}(2016){Cano-D{\'\i}az}, {S{\'a}nchez},
  {Zibetti}, {Ascasibar}, {Bland-Hawthorn}, {Ziegler}, {Gonz{\'a}lez Delgado},
  {Walcher}, {Garc{\'\i}a-Benito}, {Mast}, {Mendoza-P{\'e}rez},
  {Falc{\'o}n-Barroso}, {Galbany}, {Husemann}, {Kehrig}, {Marino},
  {S{\'a}nchez-Bl{\'a}zquez}, {L{\'o}pez-Cob{\'a}}, {L{\'o}pez-S{\'a}nchez}, \&
  {Vilchez}}]{Cano-Diaz+2016}
{Cano-D{\'\i}az}, M., {S{\'a}nchez}, S.~F., {Zibetti}, S., {et~al.} 2016,
  \apjl, 821, L26, \dodoi{10.3847/2041-8205/821/2/L26}

\bibitem[{{Caproni} {et~al.}(2015){Caproni}, {Lanfranchi}, {da Silva}, \&
  {Falceta-Gon{\c{c}}alves}}]{Caproni+2015}
{Caproni}, A., {Lanfranchi}, G.~A., {da Silva}, A.~L., \&
  {Falceta-Gon{\c{c}}alves}, D. 2015, \apj, 805, 109,
  \dodoi{10.1088/0004-637X/805/2/109}

\bibitem[{{Casey} {et~al.}(2012){Casey}, {Berta}, {B{\'e}thermin}, {Bock},
  {Bridge}, {Budynkiewicz}, {Burgarella}, {Chapin}, {Chapman}, {Clements},
  {Conley}, {Conselice}, {Cooray}, {Farrah}, {Hatziminaoglou}, {Ivison}, {le
  Floc'h}, {Lutz}, {Magdis}, {Magnelli}, {Oliver}, {Page}, {Pozzi},
  {Rigopoulou}, {Riguccini}, {Roseboom}, {Sanders}, {Scott}, {Seymour},
  {Valtchanov}, {Vieira}, {Viero}, \& {Wardlow}}]{Casey+2012}
{Casey}, C.~M., {Berta}, S., {B{\'e}thermin}, M., {et~al.} 2012, \apj, 761,
  140, \dodoi{10.1088/0004-637X/761/2/140}

\bibitem[{{Chabrier}(2003)}]{Chabrier2003}
{Chabrier}, G. 2003, \pasp, 115, 763, \dodoi{10.1086/376392}

\bibitem[{{Conselice} {et~al.}(2016){Conselice}, {Wilkinson}, {Duncan}, \&
  {Mortlock}}]{Conselice+2016}
{Conselice}, C.~J., {Wilkinson}, A., {Duncan}, K., \& {Mortlock}, A. 2016,
  \apj, 830, 83, \dodoi{10.3847/0004-637X/830/2/83}

\bibitem[{{Cucciati} {et~al.}(2012){Cucciati}, {Tresse}, {Ilbert}, {Le
  F{\`e}vre}, {Garilli}, {Le Brun}, {Cassata}, {Franzetti}, {Maccagni},
  {Scodeggio}, {Zucca}, {Zamorani}, {Bardelli}, {Bolzonella}, {Bielby},
  {McCracken}, {Zanichelli}, \& {Vergani}}]{Cucciati+2012}
{Cucciati}, O., {Tresse}, L., {Ilbert}, O., {et~al.} 2012, \aap, 539, A31,
  \dodoi{10.1051/0004-6361/201118010}

\bibitem[{{da Cunha} {et~al.}(2008){da Cunha}, {Charlot}, \&
  {Elbaz}}]{daCunha+2008}
{da Cunha}, E., {Charlot}, S., \& {Elbaz}, D. 2008, \mnras, 388, 1595,
  \dodoi{10.1111/j.1365-2966.2008.13535.x}

\bibitem[{{Daddi} {et~al.}(2007){Daddi}, {Dickinson}, {Morrison}, {Chary},
  {Cimatti}, {Elbaz}, {Frayer}, {Renzini}, {Pope}, {Alexander}, {Bauer},
  {Giavalisco}, {Huynh}, {Kurk}, \& {Mignoli}}]{Daddi+2007}
{Daddi}, E., {Dickinson}, M., {Morrison}, G., {et~al.} 2007, \apj, 670, 156,
  \dodoi{10.1086/521818}

\bibitem[{{Dalcanton} {et~al.}(2004){Dalcanton}, {Yoachim}, \&
  {Bernstein}}]{Dalcanton+2004}
{Dalcanton}, J.~J., {Yoachim}, P., \& {Bernstein}, R.~A. 2004, \apj, 608, 189,
  \dodoi{10.1086/386358}

\bibitem[{{Dav{\'e}} {et~al.}(2012){Dav{\'e}}, {Finlator}, \&
  {Oppenheimer}}]{Romeel+2012}
{Dav{\'e}}, R., {Finlator}, K., \& {Oppenheimer}, B.~D. 2012, \mnras, 421, 98,
  \dodoi{10.1111/j.1365-2966.2011.20148.x}

\bibitem[{{Davidzon} {et~al.}(2017){Davidzon}, {Ilbert}, {Laigle}, {Coupon},
  {McCracken}, {Delvecchio}, {Masters}, {Capak}, {Hsieh}, {Le F{\`e}vre},
  {Tresse}, {Bethermin}, {Chang}, {Faisst}, {Le Floc'h}, {Steinhardt}, {Toft},
  {Aussel}, {Dubois}, {Hasinger}, {Salvato}, {Sanders}, {Scoville}, \&
  {Silverman}}]{Davidzon+2017}
{Davidzon}, I., {Ilbert}, O., {Laigle}, C., {et~al.} 2017, \aap, 605, A70,
  \dodoi{10.1051/0004-6361/201730419}

\bibitem[{{Davies} {et~al.}(2019){Davies}, {Lagos}, {Katsianis}, {Robotham},
  {Cortese}, {Driver}, {Bremer}, {Brown}, {Brough}, {Cluver}, {Grootes},
  {Holwerda}, {Owers}, \& {Phillipps}}]{Davies+2019}
{Davies}, L.~J.~M., {Lagos}, C. d.~P., {Katsianis}, A., {et~al.} 2019, \mnras,
  483, 1881, \dodoi{10.1093/mnras/sty2957}

\bibitem[{{Driver} {et~al.}(2012){Driver}, {Robotham}, {Kelvin}, {Alpaslan},
  {Baldry}, {Bamford}, {Brough}, {Brown}, {Hopkins}, {Liske}, {Loveday},
  {Norberg}, {Peacock}, {Andrae}, {Bland-Hawthorn}, {Bourne}, {Cameron},
  {Colless}, {Conselice}, {Croom}, {Dunne}, {Frenk}, {Graham}, {Gunawardhana},
  {Hill}, {Jones}, {Kuijken}, {Madore}, {Nichol}, {Parkinson}, {Pimbblet},
  {Phillipps}, {Popescu}, {Prescott}, {Seibert}, {Sharp}, {Sutherland},
  {Taylor}, {Thomas}, {Tuffs}, {van Kampen}, {Wijesinghe}, \&
  {Wilkins}}]{Driver+2012}
{Driver}, S.~P., {Robotham}, A.~S.~G., {Kelvin}, L., {et~al.} 2012, \mnras,
  427, 3244, \dodoi{10.1111/j.1365-2966.2012.22036.x}

\bibitem[{{Driver} {et~al.}(2018){Driver}, {Andrews}, {da Cunha}, {Davies},
  {Lagos}, {Robotham}, {Vinsen}, {Wright}, {Alpaslan}, {Bland -Hawthorn},
  {Bourne}, {Brough}, {Bremer}, {Cluver}, {Colless}, {Conselice}, {Dunne},
  {Eales}, {Gomez}, {Holwerda}, {Hopkins}, {Kafle}, {Kelvin}, {Loveday},
  {Liske}, {Maddox}, {Phillipps}, {Pimbblet}, {Rowlands}, {Sansom}, {Taylor},
  {Wang}, \& {Wilkins}}]{Driver+2018}
{Driver}, S.~P., {Andrews}, S.~K., {da Cunha}, E., {et~al.} 2018, \mnras, 475,
  2891, \dodoi{10.1093/mnras/stx2728}

\bibitem[{{Duncan} {et~al.}(2014){Duncan}, {Conselice}, {Mortlock}, {Hartley},
  {Guo}, {Ferguson}, {Dav{\'e}}, {Lu}, {Ownsworth}, {Ashby}, {Dekel},
  {Dickinson}, {Faber}, {Giavalisco}, {Grogin}, {Kocevski}, {Koekemoer},
  {Somerville}, \& {White}}]{Duncan+2014}
{Duncan}, K., {Conselice}, C.~J., {Mortlock}, A., {et~al.} 2014, \mnras, 444,
  2960, \dodoi{10.1093/mnras/stu1622}

\bibitem[{{Dwek} \& {Cherchneff}(2011)}]{Dwek+2011}
{Dwek}, E., \& {Cherchneff}, I. 2011, \apj, 727, 63,
  \dodoi{10.1088/0004-637X/727/2/63}

\bibitem[{{Elbaz} {et~al.}(2007){Elbaz}, {Daddi}, {Le Borgne}, {Dickinson},
  {Alexander}, {Chary}, {Starck}, {Brand t}, {Kitzbichler}, {MacDonald},
  {Nonino}, {Popesso}, {Stern}, \& {Vanzella}}]{Elbaz+2007}
{Elbaz}, D., {Daddi}, E., {Le Borgne}, D., {et~al.} 2007, \aap, 468, 33,
  \dodoi{10.1051/0004-6361:20077525}

\bibitem[{{Fang} {et~al.}(2018){Fang}, {Faber}, {Koo}, {Rodr{\'\i}guez-Puebla},
  {Guo}, {Barro}, {Behroozi}, {Brammer}, {Chen}, {Dekel}, {Ferguson},
  {Gawiser}, {Giavalisco}, {Kartaltepe}, {Kocevski}, {Koekemoer}, {McGrath},
  {McIntosh}, {Newman}, {Pacifici}, {Pandya}, {P{\'e}rez-Gonz{\'a}lez},
  {Primack}, {Salmon}, {Trump}, {Weiner}, {Willner}, {Acquaviva}, {Dahlen},
  {Finkelstein}, {Finlator}, {Fontana}, {Galametz}, {Grogin}, {Gruetzbauch},
  {Johnson}, {Mobasher}, {Papovich}, {Pforr}, {Salvato}, {Santini}, {van der
  Wel}, {Wiklind}, \& {Wuyts}}]{Fang+2018}
{Fang}, J.~J., {Faber}, S.~M., {Koo}, D.~C., {et~al.} 2018, \apj, 858, 100,
  \dodoi{10.3847/1538-4357/aabcba}

\bibitem[{{Finkelstein} {et~al.}(2015){Finkelstein}, {Ryan}, {Papovich},
  {Dickinson}, {Song}, {Somerville}, {Ferguson}, {Salmon}, {Giavalisco},
  {Koekemoer}, {Ashby}, {Behroozi}, {Castellano}, {Dunlop}, {Faber}, {Fazio},
  {Fontana}, {Grogin}, {Hathi}, {Jaacks}, {Kocevski}, {Livermore}, {McLure},
  {Merlin}, {Mobasher}, {Newman}, {Rafelski}, {Tilvi}, \&
  {Willner}}]{Finkelstein+2015}
{Finkelstein}, S.~L., {Ryan}, Jr., R.~E., {Papovich}, C., {et~al.} 2015, \apj,
  810, 71, \dodoi{10.1088/0004-637X/810/1/71}

\bibitem[{{Franco} \& {Cox}(1986)}]{Franco+1986}
{Franco}, J., \& {Cox}, D.~P. 1986, \pasp, 98, 1076, \dodoi{10.1086/131876}

\bibitem[{{Gavazzi} {et~al.}(2015){Gavazzi}, {Consolandi}, {Dotti}, {Fanali},
  {Fossati}, {Fumagalli}, {Viscardi}, {Savorgnan}, {Boselli}, {Guti{\'e}rrez},
  {Hern{\'a}ndez Toledo}, {Giovanelli}, \& {Haynes}}]{Gavazzi+2015}
{Gavazzi}, G., {Consolandi}, G., {Dotti}, M., {et~al.} 2015, \aap, 580, A116,
  \dodoi{10.1051/0004-6361/201425351}

\bibitem[{{Gonz{\'a}lez} {et~al.}(2011){Gonz{\'a}lez}, {Labb{\'e}}, {Bouwens},
  {Illingworth}, {Franx}, \& {Kriek}}]{Gonzalez+2011}
{Gonz{\'a}lez}, V., {Labb{\'e}}, I., {Bouwens}, R.~J., {et~al.} 2011, \apjl,
  735, L34, \dodoi{10.1088/2041-8205/735/2/L34}

\bibitem[{{Gruppioni} {et~al.}(2013){Gruppioni}, {Pozzi}, {Rodighiero},
  {Delvecchio}, {Berta}, {Pozzetti}, {Zamorani}, {Andreani}, {Cimatti},
  {Ilbert}, {Le Floc'h}, {Lutz}, {Magnelli}, {Marchetti}, {Monaco}, {Nordon},
  {Oliver}, {Popesso}, {Riguccini}, {Roseboom}, {Rosario}, {Sargent},
  {Vaccari}, {Altieri}, {Aussel}, {Bongiovanni}, {Cepa}, {Daddi},
  {Dom{\'\i}nguez-S{\'a}nchez}, {Elbaz}, {F{\"o}rster Schreiber}, {Genzel},
  {Iribarrem}, {Magliocchetti}, {Maiolino}, {Poglitsch}, {P{\'e}rez
  Garc{\'\i}a}, {Sanchez-Portal}, {Sturm}, {Tacconi}, {Valtchanov}, {Amblard},
  {Arumugam}, {Bethermin}, {Bock}, {Boselli}, {Buat}, {Burgarella},
  {Castro-Rodr{\'\i}guez}, {Cava}, {Chanial}, {Clements}, {Conley}, {Cooray},
  {Dowell}, {Dwek}, {Eales}, {Franceschini}, {Glenn}, {Griffin},
  {Hatziminaoglou}, {Ibar}, {Isaak}, {Ivison}, {Lagache}, {Levenson}, {Lu},
  {Madden}, {Maffei}, {Mainetti}, {Nguyen}, {O'Halloran}, {Page}, {Panuzzo},
  {Papageorgiou}, {Pearson}, {P{\'e}rez-Fournon}, {Pohlen}, {Rigopoulou},
  {Rowan-Robinson}, {Schulz}, {Scott}, {Seymour}, {Shupe}, {Smith}, {Stevens},
  {Symeonidis}, {Trichas}, {Tugwell}, {Vigroux}, {Wang}, {Wright}, {Xu},
  {Zemcov}, {Bardelli}, {Carollo}, {Contini}, {Le F{\'e}vre}, {Lilly},
  {Mainieri}, {Renzini}, {Scodeggio}, \& {Zucca}}]{Gruppioni+2013}
{Gruppioni}, C., {Pozzi}, F., {Rodighiero}, G., {et~al.} 2013, \mnras, 432, 23,
  \dodoi{10.1093/mnras/stt308}

\bibitem[{{Hathi} {et~al.}(2010){Hathi}, {Ryan}, {Cohen}, {Yan}, {Windhorst},
  {McCarthy}, {O'Connell}, {Koekemoer}, {Rutkowski}, {Balick}, {Bond},
  {Calzetti}, {Disney}, {Dopita}, {Frogel}, {Hall}, {Holtzman}, {Kimble},
  {Paresce}, {Saha}, {Silk}, {Trauger}, {Walker}, {Whitmore}, \&
  {Young}}]{Hathi+2010}
{Hathi}, N.~P., {Ryan}, R.~E., J., {Cohen}, S.~H., {et~al.} 2010, \apj, 720,
  1708, \dodoi{10.1088/0004-637X/720/2/1708}

\bibitem[{{Ilbert} {et~al.}(2013){Ilbert}, {McCracken}, {Le F{\`e}vre},
  {Capak}, {Dunlop}, {Karim}, {Renzini}, {Caputi}, {Boissier}, {Arnouts},
  {Aussel}, {Comparat}, {Guo}, {Hudelot}, {Kartaltepe}, {Kneib}, {Krogager},
  {Le Floc'h}, {Lilly}, {Mellier}, {Milvang-Jensen}, {Moutard}, {Onodera},
  {Richard}, {Salvato}, {Sanders}, {Scoville}, {Silverman}, {Taniguchi},
  {Tasca}, {Thomas}, {Toft}, {Tresse}, {Vergani}, {Wolk}, \&
  {Zirm}}]{Ilbert+2013}
{Ilbert}, O., {McCracken}, H.~J., {Le F{\`e}vre}, O., {et~al.} 2013, \aap, 556,
  A55, \dodoi{10.1051/0004-6361/201321100}

\bibitem[{{Ilbert} {et~al.}(2015){Ilbert}, {Arnouts}, {Le Floc'h}, {Aussel},
  {Bethermin}, {Capak}, {Hsieh}, {Kajisawa}, {Karim}, {Le F{\`e}vre}, {Lee},
  {Lilly}, {McCracken}, {Michel-Dansac}, {Moutard}, {Renzini}, {Salvato},
  {Sanders}, {Scoville}, {Sheth}, {Silverman}, {Smol{\v c}i{\'c}}, {Taniguchi},
  \& {Tresse}}]{Ilbert+2015}
{Ilbert}, O., {Arnouts}, S., {Le Floc'h}, E., {et~al.} 2015, \aap, 579, A2,
  \dodoi{10.1051/0004-6361/201425176}

\bibitem[{{Iyer} {et~al.}(2018){Iyer}, {Gawiser}, {Dav{\'e}}, {Davis},
  {Finkelstein}, {Kodra}, {Koekemoer}, {Kurczynski}, {Newman}, {Pacifici}, \&
  {Somerville}}]{Iyer+2018}
{Iyer}, K., {Gawiser}, E., {Dav{\'e}}, R., {et~al.} 2018, \apj, 866, 120,
  \dodoi{10.3847/1538-4357/aae0fa}

\bibitem[{{Karim} {et~al.}(2011){Karim}, {Schinnerer},
  {Mart{\'{\i}}nez-Sansigre}, {Sargent}, {van der Wel}, {Rix}, {Ilbert},
  {Smol{\v c}i{\'c}}, {Carilli}, {Pannella}, {Koekemoer}, {Bell}, \&
  {Salvato}}]{Karim+2011}
{Karim}, A., {Schinnerer}, E., {Mart{\'{\i}}nez-Sansigre}, A., {et~al.} 2011,
  \apj, 730, 61, \dodoi{10.1088/0004-637X/730/2/61}

\bibitem[{{Kennicutt}(1998)}]{Kennicutt1998}
{Kennicutt}, Robert~C., J. 1998, \araa, 36, 189,
  \dodoi{10.1146/annurev.astro.36.1.189}

\bibitem[{{Khusanova} {et~al.}(2020){Khusanova}, {Le F{\`e}vre}, {Cassata},
  {Cucciati}, {Lemaux}, {Tasca}, {Thomas}, {Garilli}, {Le Brun}, {Maccagni},
  {Pentericci}, {Zamorani}, {Amor{\'\i}n}, {Bardelli}, {Castellano},
  {Cassar{\`a}}, {Cimatti}, {Giavalisco}, {Hathi}, {Ilbert}, {Koekemoer},
  {Marchi}, {Pforr}, {Ribeiro}, {Schaerer}, {Tresse}, {Vergani}, \&
  {Zucca}}]{Khusanova+2020}
{Khusanova}, Y., {Le F{\`e}vre}, O., {Cassata}, P., {et~al.} 2020, \aap, 634,
  A97, \dodoi{10.1051/0004-6361/201935400}

\bibitem[{{Kilerci Eser} \& {Goto}(2018)}]{KilerciEser+2018}
{Kilerci Eser}, E., \& {Goto}, T. 2018, \mnras, 474, 5363,
  \dodoi{10.1093/mnras/stx3110}

\bibitem[{{Kurczynski} {et~al.}(2016){Kurczynski}, {Gawiser}, {Acquaviva},
  {Bell}, {Dekel}, {de Mello}, {Ferguson}, {Gardner}, {Grogin}, {Guo},
  {Hopkins}, {Koekemoer}, {Koo}, {Lee}, {Mobasher}, {Primack}, {Rafelski},
  {Soto}, \& {Teplitz}}]{Kurczynski+2016}
{Kurczynski}, P., {Gawiser}, E., {Acquaviva}, V., {et~al.} 2016, \apjl, 820,
  L1, \dodoi{10.3847/2041-8205/820/1/L1}

\bibitem[{{Le Floc'h} {et~al.}(2005){Le Floc'h}, {Papovich}, {Dole}, {Bell},
  {Lagache}, {Rieke}, {Egami}, {P{\'e}rez-Gonz{\'a}lez}, {Alonso-Herrero},
  {Rieke}, {Blaylock}, {Engelbracht}, {Gordon}, {Hines}, {Misselt}, {Morrison},
  \& {Mould}}]{LeFloc'h+2005}
{Le Floc'h}, E., {Papovich}, C., {Dole}, H., {et~al.} 2005, \apj, 632, 169,
  \dodoi{10.1086/432789}

\bibitem[{{Lee} {et~al.}(2018){Lee}, {Giavalisco}, {Whitaker}, {Williams},
  {Ferguson}, {Acquaviva}, {Koekemoer}, {Straughn}, {Guo}, {Kartaltepe},
  {Lotz}, {Pacifici}, {Croton}, {Somerville}, \& {Lu}}]{Lee+2018}
{Lee}, B., {Giavalisco}, M., {Whitaker}, K., {et~al.} 2018, \apj, 853, 131,
  \dodoi{10.3847/1538-4357/aaa40f}

\bibitem[{{Lee} {et~al.}(2015){Lee}, {Sanders}, {Casey}, {Toft}, {Scoville},
  {Hung}, {Le Floc'h}, {Ilbert}, {Zahid}, {Aussel}, {Capak}, {Kartaltepe},
  {Kewley}, {Li}, {Schawinski}, {Sheth}, \& {Xiao}}]{Lee+2015}
{Lee}, N., {Sanders}, D.~B., {Casey}, C.~M., {et~al.} 2015, \apj, 801, 80,
  \dodoi{10.1088/0004-637X/801/2/80}

\bibitem[{{Leslie} {et~al.}(2020){Leslie}, {Schinnerer}, {Liu}, {Magnelli},
  {Algera}, {Karim}, {Davidzon}, {Gozaliasl}, {Jim{\'e}nez-Andrade}, {Lang},
  {Sargent}, {Novak}, {Groves}, {Smol{\v{c}}i{\'c}}, {Zamorani}, {Vaccari},
  {Battisti}, {Vardoulaki}, {Peng}, \& {Kartaltepe}}]{Leslie+2020}
{Leslie}, S.~K., {Schinnerer}, E., {Liu}, D., {et~al.} 2020, \apj, 899, 58,
  \dodoi{10.3847/1538-4357/aba044}

\bibitem[{{Lim} {et~al.}(2020){Lim}, {Wang}, {Smail}, {Scott}, {Chen}, {Chang},
  {Simpson}, {Toba}, {Shu}, {Clements}, {Greenslade}, {Ao}, {Babul}, {Birkin},
  {Chapman}, {Cheng}, {Cho}, {Dannerbauer}, {Dudzevi{\v{c}}i{\={u}}t{\.{e}}},
  {Dunlop}, {Gao}, {Goto}, {Ho}, {Hsu}, {Hwang}, {Jeong}, {Koprowski}, {Lee},
  {Lin}, {Lin}, {Micha{\l}owski}, {Parsons}, {Sawicki}, {Shirley}, {Shim},
  {Urquhart}, {Wang}, \& {Wang}}]{Lim+2020}
{Lim}, C.-F., {Wang}, W.-H., {Smail}, I., {et~al.} 2020, \apj, 889, 80,
  \dodoi{10.3847/1538-4357/ab607f}

\bibitem[{{Liu} {et~al.}(2018){Liu}, {Daddi}, {Dickinson}, {Owen}, {Pannella},
  {Sargent}, {B{\'e}thermin}, {Magdis}, {Gao}, {Shu}, {Wang}, {Jin}, \&
  {Inami}}]{Liu+2018}
{Liu}, D., {Daddi}, E., {Dickinson}, M., {et~al.} 2018, \apj, 853, 172,
  \dodoi{10.3847/1538-4357/aaa600}

\bibitem[{{Livermore} {et~al.}(2017){Livermore}, {Finkelstein}, \&
  {Lotz}}]{Livermore+2017}
{Livermore}, R.~C., {Finkelstein}, S.~L., \& {Lotz}, J.~M. 2017, \apj, 835,
  113, \dodoi{10.3847/1538-4357/835/2/113}

\bibitem[{{Madau} \& {Dickinson}(2014)}]{Madau_Dickinson2014}
{Madau}, P., \& {Dickinson}, M. 2014, \araa, 52, 415,
  \dodoi{10.1146/annurev-astro-081811-125615}

\bibitem[{{Magnelli} {et~al.}(2011){Magnelli}, {Elbaz}, {Chary}, {Dickinson},
  {Le Borgne}, {Frayer}, \& {Willmer}}]{Magnelli+2011}
{Magnelli}, B., {Elbaz}, D., {Chary}, R.~R., {et~al.} 2011, \aap, 528, A35,
  \dodoi{10.1051/0004-6361/200913941}

\bibitem[{{Magnelli} {et~al.}(2013){Magnelli}, {Popesso}, {Berta}, {Pozzi},
  {Elbaz}, {Lutz}, {Dickinson}, {Altieri}, {Andreani}, {Aussel},
  {B{\'e}thermin}, {Bongiovanni}, {Cepa}, {Charmandaris}, {Chary}, {Cimatti},
  {Daddi}, {F{\"o}rster Schreiber}, {Genzel}, {Gruppioni}, {Harwit}, {Hwang},
  {Ivison}, {Magdis}, {Maiolino}, {Murphy}, {Nordon}, {Pannella}, {P{\'e}rez
  Garc{\'\i}a}, {Poglitsch}, {Rosario}, {Sanchez-Portal}, {Santini}, {Scott},
  {Sturm}, {Tacconi}, \& {Valtchanov}}]{Magnelli+2013}
{Magnelli}, B., {Popesso}, P., {Berta}, S., {et~al.} 2013, \aap, 553, A132,
  \dodoi{10.1051/0004-6361/201321371}

\bibitem[{{McGaugh} {et~al.}(2017){McGaugh}, {Schombert}, \&
  {Lelli}}]{McGaugh+2017}
{McGaugh}, S.~S., {Schombert}, J.~M., \& {Lelli}, F. 2017, \apj, 851, 22,
  \dodoi{10.3847/1538-4357/aa9790}

\bibitem[{{Mehta} {et~al.}(2017){Mehta}, {Scarlata}, {Rafelski}, {Gburek},
  {Teplitz}, {Alavi}, {Boylan-Kolchin}, {Finkelstein}, {Gardner}, {Grogin},
  {Koekemoer}, {Kurczynski}, {Siana}, {Codoreanu}, {de Mello}, {Lee}, \&
  {Soto}}]{Mehta+2017}
{Mehta}, V., {Scarlata}, C., {Rafelski}, M., {et~al.} 2017, \apj, 838, 29,
  \dodoi{10.3847/1538-4357/aa6259}

\bibitem[{{Moster} {et~al.}(2018){Moster}, {Naab}, \& {White}}]{Moster+2018}
{Moster}, B.~P., {Naab}, T., \& {White}, S. D.~M. 2018, \mnras, 477, 1822,
  \dodoi{10.1093/mnras/sty655}

\bibitem[{{Murphy} {et~al.}(2011){Murphy}, {Condon}, {Schinnerer}, {Kennicutt},
  {Calzetti}, {Armus}, {Helou}, {Turner}, {Aniano}, {Beir{\~a}o}, {Bolatto},
  {Brandl}, {Croxall}, {Dale}, {Donovan Meyer}, {Draine}, {Engelbracht},
  {Hunt}, {Hao}, {Koda}, {Roussel}, {Skibba}, \& {Smith}}]{Murphy+2011}
{Murphy}, E.~J., {Condon}, J.~J., {Schinnerer}, E., {et~al.} 2011, \apj, 737,
  67, \dodoi{10.1088/0004-637X/737/2/67}

\bibitem[{{Muzzin} {et~al.}(2013){Muzzin}, {Marchesini}, {Stefanon}, {Franx},
  {McCracken}, {Milvang-Jensen}, {Dunlop}, {Fynbo}, {Brammer}, {Labb{\'e}}, \&
  {van Dokkum}}]{Muzzin+2013}
{Muzzin}, A., {Marchesini}, D., {Stefanon}, M., {et~al.} 2013, \apj, 777, 18,
  \dodoi{10.1088/0004-637X/777/1/18}

\bibitem[{{Noeske} {et~al.}(2007){Noeske}, {Weiner}, {Faber}, {Papovich},
  {Koo}, {Somerville}, {Bundy}, {Conselice}, {Newman}, {Schiminovich}, {Le
  Floc'h}, {Coil}, {Rieke}, {Lotz}, {Primack}, {Barmby}, {Cooper}, {Davis},
  {Ellis}, {Fazio}, {Guhathakurta}, {Huang}, {Kassin}, {Martin}, {Phillips},
  {Rich}, {Small}, {Willmer}, \& {Wilson}}]{Noeske+2007}
{Noeske}, K.~G., {Weiner}, B.~J., {Faber}, S.~M., {et~al.} 2007, \apjl, 660,
  L43, \dodoi{10.1086/517926}

\bibitem[{{Novak} {et~al.}(2017){Novak}, {Smol{\v{c}}i{\'c}}, {Delhaize},
  {Delvecchio}, {Zamorani}, {Baran}, {Bondi}, {Capak}, {Carilli}, {Ciliegi},
  {Civano}, {Ilbert}, {Karim}, {Laigle}, {Le F{\`e}vre}, {Marchesi},
  {McCracken}, {Miettinen}, {Salvato}, {Sargent}, {Schinnerer}, \&
  {Tasca}}]{Novak+2017}
{Novak}, M., {Smol{\v{c}}i{\'c}}, V., {Delhaize}, J., {et~al.} 2017, \aap, 602,
  A5, \dodoi{10.1051/0004-6361/201629436}

\bibitem[{{Oesch} {et~al.}(2010){Oesch}, {Bouwens}, {Carollo}, {Illingworth},
  {Magee}, {Trenti}, {Stiavelli}, {Franx}, {Labb{\'e}}, \& {van
  Dokkum}}]{Oesch+2010}
{Oesch}, P.~A., {Bouwens}, R.~J., {Carollo}, C.~M., {et~al.} 2010, \apjl, 725,
  L150, \dodoi{10.1088/2041-8205/725/2/L150}

\bibitem[{{Oesch} {et~al.}(2012){Oesch}, {Bouwens}, {Illingworth}, {Gonzalez},
  {Trenti}, {van Dokkum}, {Franx}, {Labb{\'e}}, {Carollo}, \&
  {Magee}}]{Oesch+2012}
{Oesch}, P.~A., {Bouwens}, R.~J., {Illingworth}, G.~D., {et~al.} 2012, \apj,
  759, 135, \dodoi{10.1088/0004-637X/759/2/135}

\bibitem[{{Ono} {et~al.}(2018){Ono}, {Ouchi}, {Harikane}, {Toshikawa}, {Rauch},
  {Yuma}, {Sawicki}, {Shibuya}, {Shimasaku}, {Oguri}, {Willott}, {Akhlaghi},
  {Akiyama}, {Coupon}, {Kashikawa}, {Komiyama}, {Konno}, {Lin}, {Matsuoka},
  {Miyazaki}, {Nagao}, {Nakajima}, {Silverman}, {Tanaka}, {Taniguchi}, \&
  {Wang}}]{Ono+2018}
{Ono}, Y., {Ouchi}, M., {Harikane}, Y., {et~al.} 2018, \pasj, 70, S10,
  \dodoi{10.1093/pasj/psx103}

\bibitem[{{Pannella} {et~al.}(2009){Pannella}, {Carilli}, {Daddi}, {McCracken},
  {Owen}, {Renzini}, {Strazzullo}, {Civano}, {Koekemoer}, {Schinnerer},
  {Scoville}, {Smol{\v{c}}i{\'c}}, {Taniguchi}, {Aussel}, {Kneib}, {Ilbert},
  {Mellier}, {Salvato}, {Thompson}, \& {Willott}}]{Pannella+2009}
{Pannella}, M., {Carilli}, C.~L., {Daddi}, E., {et~al.} 2009, \apjl, 698, L116,
  \dodoi{10.1088/0004-637X/698/2/L116}

\bibitem[{{Parsa} {et~al.}(2016){Parsa}, {Dunlop}, {McLure}, \&
  {Mortlock}}]{Parsa+2016}
{Parsa}, S., {Dunlop}, J.~S., {McLure}, R.~J., \& {Mortlock}, A. 2016, \mnras,
  456, 3194, \dodoi{10.1093/mnras/stv2857}

\bibitem[{{Popesso} {et~al.}(2019{\natexlab{a}}){Popesso}, {Morselli},
  {Concas}, {Schreiber}, {Rodighiero}, {Cresci}, {Belli}, {Ilbert},
  {Erfanianfar}, {Mancini}, {Inami}, {Dickinson}, {Pannella}, \&
  {Elbaz}}]{Popesso+2019}
{Popesso}, P., {Morselli}, L., {Concas}, A., {et~al.} 2019{\natexlab{a}},
  \mnras, 490, 5285, \dodoi{10.1093/mnras/stz2635}

\bibitem[{{Popesso} {et~al.}(2019{\natexlab{b}}){Popesso}, {Concas},
  {Morselli}, {Schreiber}, {Rodighiero}, {Cresci}, {Belli}, {Erfanianfar},
  {Mancini}, {Inami}, {Dickinson}, {Ilbert}, {Pannella}, \&
  {Elbaz}}]{Popesso+2019a}
{Popesso}, P., {Concas}, A., {Morselli}, L., {et~al.} 2019{\natexlab{b}},
  \mnras, 483, 3213, \dodoi{10.1093/mnras/sty3210}

\bibitem[{{Reddy} {et~al.}(2012){Reddy}, {Pettini}, {Steidel}, {Shapley},
  {Erb}, \& {Law}}]{Reddy+2012}
{Reddy}, N.~A., {Pettini}, M., {Steidel}, C.~C., {et~al.} 2012, \apj, 754, 25,
  \dodoi{10.1088/0004-637X/754/1/25}

\bibitem[{{Reddy} \& {Steidel}(2009)}]{Reddy+2009}
{Reddy}, N.~A., \& {Steidel}, C.~C. 2009, \apj, 692, 778,
  \dodoi{10.1088/0004-637X/692/1/778}

\bibitem[{{Robotham} \& {Driver}(2011)}]{Robotham_Driver2011}
{Robotham}, A.~S.~G., \& {Driver}, S.~P. 2011, \mnras, 413, 2570,
  \dodoi{10.1111/j.1365-2966.2011.18327.x}

\bibitem[{{Rodighiero} {et~al.}(2010){Rodighiero}, {Vaccari}, {Franceschini},
  {Tresse}, {Le Fevre}, {Le Brun}, {Mancini}, {Matute}, {Cimatti}, {Marchetti},
  {Ilbert}, {Arnouts}, {Bolzonella}, {Zucca}, {Bardelli}, {Lonsdale}, {Shupe},
  {Surace}, {Rowan-Robinson}, {Garilli}, {Zamorani}, {Pozzetti}, {Bondi}, {de
  la Torre}, {Vergani}, {Santini}, {Grazian}, \& {Fontana}}]{Rodighiero+2010b}
{Rodighiero}, G., {Vaccari}, M., {Franceschini}, A., {et~al.} 2010, \aap, 515,
  A8, \dodoi{10.1051/0004-6361/200912058}

\bibitem[{{Rodighiero} {et~al.}(2011){Rodighiero}, {Daddi}, {Baronchelli},
  {Cimatti}, {Renzini}, {Aussel}, {Popesso}, {Lutz}, {Andreani}, {Berta},
  {Cava}, {Elbaz}, {Feltre}, {Fontana}, {F{\"o}rster Schreiber},
  {Franceschini}, {Genzel}, {Grazian}, {Gruppioni}, {Ilbert}, {Le Floch},
  {Magdis}, {Magliocchetti}, {Magnelli}, {Maiolino}, {McCracken}, {Nordon},
  {Poglitsch}, {Santini}, {Pozzi}, {Riguccini}, {Tacconi}, {Wuyts}, \&
  {Zamorani}}]{Rodighiero+2011}
{Rodighiero}, G., {Daddi}, E., {Baronchelli}, I., {et~al.} 2011, \apjl, 739,
  L40, \dodoi{10.1088/2041-8205/739/2/L40}

\bibitem[{{Rodr{\'{\i}}guez-Puebla} {et~al.}(2013){Rodr{\'{\i}}guez-Puebla},
  {Avila-Reese}, \& {Drory}}]{Rodriguez-Puebla+2013}
{Rodr{\'{\i}}guez-Puebla}, A., {Avila-Reese}, V., \& {Drory}, N. 2013, \apj,
  767, 92, \dodoi{10.1088/0004-637X/767/1/92}

\bibitem[{{Rodr{\'\i}guez-Puebla} {et~al.}(2020){Rodr{\'\i}guez-Puebla},
  {Calette}, {Avila-Reese}, {Rodriguez-Gomez}, \& {Huertas-Company}}]{RP20}
{Rodr{\'\i}guez-Puebla}, A., {Calette}, A.~R., {Avila-Reese}, V.,
  {Rodriguez-Gomez}, V., \& {Huertas-Company}, M. 2020, \pasa, 37, e024,
  \dodoi{10.1017/pasa.2020.15}

\bibitem[{{Rodr{\'{\i}}guez-Puebla} {et~al.}(2017){Rodr{\'{\i}}guez-Puebla},
  {Primack}, {Avila-Reese}, \& {Faber}}]{Rodriguez-Puebla+2017}
{Rodr{\'{\i}}guez-Puebla}, A., {Primack}, J.~R., {Avila-Reese}, V., \& {Faber},
  S.~M. 2017, \mnras, 470, 651, \dodoi{10.1093/mnras/stx1172}

\bibitem[{{Salim} {et~al.}(2018){Salim}, {Boquien}, \& {Lee}}]{Salim+2018}
{Salim}, S., {Boquien}, M., \& {Lee}, J.~C. 2018, \apj, 859, 11,
  \dodoi{10.3847/1538-4357/aabf3c}

\bibitem[{{Salim} {et~al.}(2007){Salim}, {Rich}, {Charlot}, {Brinchmann},
  {Johnson}, {Schiminovich}, {Seibert}, {Mallery}, {Heckman}, {Forster},
  {Friedman}, {Martin}, {Morrissey}, {Neff}, {Small}, {Wyder}, {Bianchi},
  {Donas}, {Lee}, {Madore}, {Milliard}, {Szalay}, {Welsh}, \&
  {Yi}}]{Salim+2007}
{Salim}, S., {Rich}, R.~M., {Charlot}, S., {et~al.} 2007, \apjs, 173, 267,
  \dodoi{10.1086/519218}

\bibitem[{{Salim} {et~al.}(2016){Salim}, {Lee}, {Janowiecki}, {da Cunha},
  {Dickinson}, {Boquien}, {Burgarella}, {Salzer}, \& {Charlot}}]{Salim+2016}
{Salim}, S., {Lee}, J.~C., {Janowiecki}, S., {et~al.} 2016, \apjs, 227, 2,
  \dodoi{10.3847/0067-0049/227/1/2}

\bibitem[{{Salmon} {et~al.}(2015){Salmon}, {Papovich}, {Finkelstein}, {Tilvi},
  {Finlator}, {Behroozi}, {Dahlen}, {Dav{\'e}}, {Dekel}, {Dickinson},
  {Ferguson}, {Giavalisco}, {Long}, {Lu}, {Mobasher}, {Reddy}, {Somerville}, \&
  {Wechsler}}]{Salmon+2015}
{Salmon}, B., {Papovich}, C., {Finkelstein}, S.~L., {et~al.} 2015, \apj, 799,
  183, \dodoi{10.1088/0004-637X/799/2/183}

\bibitem[{{Salpeter}(1955)}]{Salpeter1955}
{Salpeter}, E.~E. 1955, \apj, 121, 161, \dodoi{10.1086/145971}

\bibitem[{{S{\'a}nchez} {et~al.}(2016){S{\'a}nchez}, {P{\'e}rez},
  {S{\'a}nchez-Bl{\'a}zquez}, {Garc{\'\i}a-Benito}, {Ibarra-Mede},
  {Gonz{\'a}lez}, {Rosales-Ortega}, {S{\'a}nchez-Menguiano}, {Ascasibar},
  {Bitsakis}, {Law}, {Cano-D{\'\i}az}, {L{\'o}pez-Cob{\'a}}, {Marino}, {Gil de
  Paz}, {L{\'o}pez-S{\'a}nchez}, {Barrera-Ballesteros}, {Galbany}, {Mast},
  {Abril-Melgarejo}, \& {Roman-Lopes}}]{Sanchez+2016}
{S{\'a}nchez}, S.~F., {P{\'e}rez}, E., {S{\'a}nchez-Bl{\'a}zquez}, P., {et~al.}
  2016, \rmxaa, 52, 171.
\newblock \doarXiv{1602.01830}

\bibitem[{{S{\'a}nchez} {et~al.}(2018){S{\'a}nchez}, {Avila-Reese},
  {Hernandez-Toledo}, {Cortes-Su{\'a}rez}, {Rodr{\'\i}guez-Puebla},
  {Ibarra-Medel}, {Cano-D{\'\i}az}, {Barrera-Ballesteros}, {Negrete},
  {Calette}, {de Lorenzo-C{\'a}ceres}, {Ortega-Minakata}, {Aquino},
  {Valenzuela}, {Clemente}, {Storchi-Bergmann}, {Riffel}, {Schimoia}, {Riffel},
  {Rembold}, {Brownstein}, {Pan}, {Yates}, {Mallmann}, \&
  {Bitsakis}}]{Sanchez+2018}
{S{\'a}nchez}, S.~F., {Avila-Reese}, V., {Hernandez-Toledo}, H., {et~al.} 2018,
  \rmxaa, 54, 217.
\newblock \doarXiv{1709.05438}

\bibitem[{{Santini} {et~al.}(2017){Santini}, {Fontana}, {Castellano}, {Di
  Criscienzo}, {Merlin}, {Amorin}, {Cullen}, {Daddi}, {Dickinson}, {Dunlop},
  {Grazian}, {Lamastra}, {McLure}, {Micha{\l}owski}, {Pentericci}, \&
  {Shu}}]{Santini+2017}
{Santini}, P., {Fontana}, A., {Castellano}, M., {et~al.} 2017, \apj, 847, 76,
  \dodoi{10.3847/1538-4357/aa8874}

\bibitem[{{Sargent} {et~al.}(2012){Sargent}, {B{\'e}thermin}, {Daddi}, \&
  {Elbaz}}]{Sargent+2012}
{Sargent}, M.~T., {B{\'e}thermin}, M., {Daddi}, E., \& {Elbaz}, D. 2012, \apjl,
  747, L31, \dodoi{10.1088/2041-8205/747/2/L31}

\bibitem[{{Schreiber} {et~al.}(2015){Schreiber}, {Pannella}, {Elbaz},
  {B{\'e}thermin}, {Inami}, {Dickinson}, {Magnelli}, {Wang}, {Aussel}, {Daddi},
  {Juneau}, {Shu}, {Sargent}, {Buat}, {Faber}, {Ferguson}, {Giavalisco},
  {Koekemoer}, {Magdis}, {Morrison}, {Papovich}, {Santini}, \&
  {Scott}}]{Schreiber+2015}
{Schreiber}, C., {Pannella}, M., {Elbaz}, D., {et~al.} 2015, \aap, 575, A74,
  \dodoi{10.1051/0004-6361/201425017}

\bibitem[{{Shin} {et~al.}(2019){Shin}, {Ly}, {Malkan}, {Malhotra}, {de los
  Reyes}, \& {Rhoads}}]{Shin+2019}
{Shin}, K., {Ly}, C., {Malkan}, M.~A., {et~al.} 2019, arXiv e-prints,
  arXiv:1910.10735.
\newblock \doarXiv{1910.10735}

\bibitem[{{Skibba} {et~al.}(2011){Skibba}, {Engelbracht}, {Dale}, {Hinz},
  {Zibetti}, {Crocker}, {Groves}, {Hunt}, {Johnson}, {Meidt}, {Murphy},
  {Appleton}, {Armus}, {Bolatto}, {Brandl}, {Calzetti}, {Croxall}, {Galametz},
  {Gordon}, {Kennicutt}, {Koda}, {Krause}, {Montiel}, {Rix}, {Roussel},
  {Sandstrom}, {Sauvage}, {Schinnerer}, {Smith}, {Walter}, {Wilson}, \&
  {Wolfire}}]{Skibba+2011}
{Skibba}, R.~A., {Engelbracht}, C.~W., {Dale}, D., {et~al.} 2011, \apj, 738,
  89, \dodoi{10.1088/0004-637X/738/1/89}

\bibitem[{{Slavin} {et~al.}(2020){Slavin}, {Dwek}, {Mac Low}, \&
  {Hill}}]{Slavin+2020}
{Slavin}, J.~D., {Dwek}, E., {Mac Low}, M.-M., \& {Hill}, A.~S. 2020, arXiv
  e-prints, arXiv:2009.01895.
\newblock \doarXiv{2009.01895}

\bibitem[{{Speagle} {et~al.}(2014){Speagle}, {Steinhardt}, {Capak}, \&
  {Silverman}}]{Speagle+2014}
{Speagle}, J.~S., {Steinhardt}, C.~L., {Capak}, P.~L., \& {Silverman}, J.~D.
  2014, \apjs, 214, 15, \dodoi{10.1088/0067-0049/214/2/15}

\bibitem[{{Tacchella} {et~al.}(2018){Tacchella}, {Bose}, {Conroy},
  {Eisenstein}, \& {Johnson}}]{Tacchella+2018}
{Tacchella}, S., {Bose}, S., {Conroy}, C., {Eisenstein}, D.~J., \& {Johnson},
  B.~D. 2018, \apj, 868, 92, \dodoi{10.3847/1538-4357/aae8e0}

\bibitem[{{Tasca} {et~al.}(2015){Tasca}, {Le F{\`e}vre}, {Hathi}, {Schaerer},
  {Ilbert}, {Zamorani}, {Lemaux}, {Cassata}, {Garilli}, {Le Brun}, {Maccagni},
  {Pentericci}, {Thomas}, {Vanzella}, {Zucca}, {Amorin}, {Bardelli},
  {Cassar{\`a}}, {Castellano}, {Cimatti}, {Cucciati}, {Durkalec}, {Fontana},
  {Giavalisco}, {Grazian}, {Paltani}, {Ribeiro}, {Scodeggio}, {Sommariva},
  {Talia}, {Tresse}, {Vergani}, {Capak}, {Charlot}, {Contini}, {de la Torre},
  {Dunlop}, {Fotopoulou}, {Koekemoer}, {L{\'o}pez-Sanjuan}, {Mellier}, {Pforr},
  {Salvato}, {Scoville}, {Taniguchi}, \& {Wang}}]{Tasca+2015}
{Tasca}, L.~A.~M., {Le F{\`e}vre}, O., {Hathi}, N.~P., {et~al.} 2015, \aap,
  581, A54, \dodoi{10.1051/0004-6361/201425379}

\bibitem[{{Tomczak} {et~al.}(2014){Tomczak}, {Quadri}, {Tran}, {Labb{\'e}},
  {Straatman}, {Papovich}, {Glazebrook}, {Allen}, {Brammer}, {Kacprzak},
  {Kawinwanichakij}, {Kelson}, {McCarthy}, {Mehrtens}, {Monson}, {Persson},
  {Spitler}, {Tilvi}, \& {van Dokkum}}]{Tomczak+2014}
{Tomczak}, A.~R., {Quadri}, R.~F., {Tran}, K.-V.~H., {et~al.} 2014, \apj, 783,
  85, \dodoi{10.1088/0004-637X/783/2/85}

\bibitem[{{Tomczak} {et~al.}(2016){Tomczak}, {Quadri}, {Tran}, {Labb{\'e}},
  {Straatman}, {Papovich}, {Glazebrook}, {Allen}, {Brammer}, {Cowley},
  {Dickinson}, {Elbaz}, {Inami}, {Kacprzak}, {Morrison}, {Nanayakkara},
  {Persson}, {Rees}, {Salmon}, {Schreiber}, {Spitler}, \&
  {Whitaker}}]{Tomczak+2016}
---. 2016, \apj, 817, 118, \dodoi{10.3847/0004-637X/817/2/118}

\bibitem[{{van der Burg} {et~al.}(2010){van der Burg}, {Hildebrandt}, \&
  {Erben}}]{vanderBurg+2010}
{van der Burg}, R.~F.~J., {Hildebrandt}, H., \& {Erben}, T. 2010, \aap, 523,
  A74, \dodoi{10.1051/0004-6361/200913812}

\bibitem[{{Weisz} {et~al.}(2014){Weisz}, {Dolphin}, {Skillman}, {Holtzman},
  {Gilbert}, {Dalcanton}, \& {Williams}}]{Weisz+2014}
{Weisz}, D.~R., {Dolphin}, A.~E., {Skillman}, E.~D., {et~al.} 2014, \apj, 789,
  147, \dodoi{10.1088/0004-637X/789/2/147}

\bibitem[{{Whitaker} {et~al.}(2017){Whitaker}, {Pope}, {Cybulski}, {Casey},
  {Popping}, \& {Yun}}]{Whitaker+2017}
{Whitaker}, K.~E., {Pope}, A., {Cybulski}, R., {et~al.} 2017, \apj, 850, 208,
  \dodoi{10.3847/1538-4357/aa94ce}

\bibitem[{{Whitaker} {et~al.}(2012){Whitaker}, {van Dokkum}, {Brammer}, \&
  {Franx}}]{Whitaker+2012}
{Whitaker}, K.~E., {van Dokkum}, P.~G., {Brammer}, G., \& {Franx}, M. 2012,
  \apjl, 754, L29, \dodoi{10.1088/2041-8205/754/2/L29}

\bibitem[{{Whitaker} {et~al.}(2014){Whitaker}, {Franx}, {Leja}, {van Dokkum},
  {Henry}, {Skelton}, {Fumagalli}, {Momcheva}, {Brammer}, {Labb{\'e}},
  {Nelson}, \& {Rigby}}]{Whitaker+2014}
{Whitaker}, K.~E., {Franx}, M., {Leja}, J., {et~al.} 2014, \apj, 795, 104,
  \dodoi{10.1088/0004-637X/795/2/104}

\bibitem[{{Williams} {et~al.}(2009){Williams}, {Quadri}, {Franx}, {van Dokkum},
  \& {Labb{\'e}}}]{Williams+2009}
{Williams}, R.~J., {Quadri}, R.~F., {Franx}, M., {van Dokkum}, P., \&
  {Labb{\'e}}, I. 2009, \apj, 691, 1879, \dodoi{10.1088/0004-637X/691/2/1879}

\bibitem[{{Wuyts} {et~al.}(2007){Wuyts}, {Labb{\'e}}, {Franx}, {Rudnick}, {van
  Dokkum}, {Fazio}, {F{\"o}rster Schreiber}, {Huang}, {Moorwood}, {Rix},
  {R{\"o}ttgering}, \& {van der Werf}}]{Wuyts+2007}
{Wuyts}, S., {Labb{\'e}}, I., {Franx}, M., {et~al.} 2007, \apj, 655, 51,
  \dodoi{10.1086/509708}

\end{thebibliography}
\bibliographystyle{aasjournal}



\end{document}